\documentclass[a4paper,fleqn,usenatbib,useAMS]{mnras}

\usepackage{graphicx,multicol} 
\usepackage{times}
\def\etal{\it et~al.}

\title[Drifting, Nulling \& Mode changing in PSR J2006$-$0807]{Subpulse Drifting, Nulling and Mode changing in PSR J2006$-$0807 with Core emission}
\author[Basu, Paul \& Mitra]{Rahul Basu$^{1,2}$, Ashis Paul$^{3}$, Dipanjan Mitra$^{3,2}$ \\
$^{1}$ Inter-University Centre for Astronomy and Astrophysics, Pune, 411007, India; rahulbasu.astro@gmail.com \\
$^{2}$ Janusz Gil Institute of Astronomy, University of Zielona G\'ora, ul. Szafrana 2, 65-516 Zielona G\'ora, Poland \\
$^{3}$ National Centre for Radio Astrophysics, Tata Institute of Fundamental Research, Pune 411007, India \\
}
\begin{document}



\maketitle

\label{firstpage}

\begin{abstract}
We report a detailed analysis of the emission behaviour of the five component, 
core-double cone, pulsar J2006$-$0807 (B2003$-$08). The single pulses 
revealed the presence of the three major phenomena of subpulse drifting, 
nulling and mode changing. The pulsar switched between four different emission 
modes, two of which showed systematic drifting with prominent drift bands, and 
were classified as modes A and B respectively. The drifting was seen primarily 
in the conal components and exhibited the rare phenomenon of bi-drifting, where 
the drift direction in the second component was opposite to the fourth 
component. This made PSR J2006$-$0807 the only known example where systematic 
drift bands were seen around a central core emission. The emission showed a 
gradual decrease in intensity during mode A which stabilised to a relatively 
constant level in the subsequent mode B. The presence of a low frequency, weak 
and wide structure in the fluctuation spectra was also seen primarily in the 
core component during modes A and B. The core component vanished during mode C 
and was most prominent during the fourth mode D. Both these modes were 
frequently interspersed with null pulses. No detectable drifting was seen 
during modes C and D, but the pulsar showed short duration periodic nulling in 
the core as well as the conal components. In addition to the four emission 
modes the pulsar also nulled for long durations lasting up to hundred rotation 
periods.
\end{abstract}


\section{Introduction}
\noindent
The single pulse sequences in the radio emission from certain pulsars are 
characterised by the three prominent phenomena of nulling, mode changing and 
subpulse drifting. The nulling and mode changing are seen as sudden changes in 
the emission state within a few rotation periods. The emission changes to a 
different profile shape in case of mode changing, while during nulling the 
radio emission completely ceases for certain durations. The subpulse drifting 
on the other hand is seen as a periodic pattern within the pulse window where 
the subpulses show gradual shift resembling drift bands. The average profile is
expected to be representative of the radio emission region of the 
magnetosphere, which is approximated as an emission beam. The emission beam 
comprises of a central core component, surrounded by nested rings of conal 
emission \citep{ran93,mit99,mit02}. Recent studies have shown the core and 
conal components to originate from similar emission heights 
\citep{mac12,skr18}. The shape of the average profile depends on the line of 
sight (LOS) traverse of the emission beam. The profile shape can be categorised
as conal single (S$_d$), conal Double (D), conal Triple ($_c$T), conal 
Quadruple ($_c$Q), core-cone Triple (T) and core-double cone Multiple (M) with 
five components. This corresponds to the different LOS traverses progressively 
from the edge of the emission beam towards the center. There are also the core 
single (S$_t$) profiles, with central LOS traverse, where the conal emission 
are too weak to be detected at frequencies below 1 GHz \citep{mit17}. A number 
of pulsars have now been reported where the pulsar transitions between 
different emission modes with some of them exhibiting drifting. The examples of
such pulsars are B0031$-$07 \citep{viv97,smi05,mcs17}, B1918+19 \citep{han87,
ran13}, B1944+17 \citep{klo10}, J1822$-$2256 \citep{bas18b} and B2319+60 
\citep{wri81}. These pulsars show a wide variety in their profile shapes but 
they all have conal profiles.

There are very few studies which report similar behaviour in core-cone pulsars,
particularly with M type profiles, with B1237+25 one of the few well studied 
examples of a five component pulsar showing presence of subpulse drifting as
well as mode changing \citep{bac70,sro05,smi13,maa14}. The presence of subpulse
drifting and two emission modes has also been reported for the pulsar B1737+13,
which has five components with a central core emission \citep{for10}. However, 
in none of these cases the pulsar shows multiple emission modes with systematic
drift bands and nulling. This has led to a number of generalized assumptions 
about the radio emission in the presence of core component. For example, it has
been suggested that the drifting is primarily phase stationary for the central 
line of sight traverse of the emission beam, corresponding to T and M profiles 
\citep{ran86,gil00}. The systematic drift bands are only possible in conal 
profiles with peripheral line of sight traverses of the emission beam. 
Additionally, the appearance of periodic nulling in certain pulsars has been 
seen in primarily conal profiles. An association between periodic nulls and 
drifting has been proposed, with the periodic nulls also termed as 
pseudo-nulls. They are postulated to originate due to the line of sight 
traverses between empty regions of a rotating sub-beam system \citep{her07,
her09}. Detailed observations of pulsars with core emission in recent works 
suggest that revisions of our prevalent understanding of the emission processes
are necessary. \citet{bas19} has studied the phase behaviour of drifting in 
pulsars with core components and found significant phase variations in the 
corresponding conal components. It has also been shown by \citet{bas16,bas17} 
that the drifting periodicity is inversely proportional to the spin-down energy
loss ($\dot{E}$), while the periodicity associated with nulling show no such 
dependence. This suggests different physical mechanisms responsible for the two
phenomenon. However, the biggest challenge has been to find a pulsar with core 
emission which show clear evidence of drift bands in addition to multiple 
emission modes and nulling in the same system. This would enable us to directly
study the emission processes for central line of sight traverse of the emission
beam. 

The pulsar J2006$-$0807 (B2003$-$08) has a prominent core component with 
clearly resolved inner and outer cones, and belongs to profile class M. The 
pulsar was observed as part of the Meterwavelength Single-pulse Polarimetric 
Emission Survey \citep[MSPES,][]{mit16}. The pulsar exhibited subpulse drifting
whose average behaviour was reported in \citet{bas16}. In addition, the single 
pulses indicated the presence of nulling as well as mode changing. The pulsar 
has the potential to be an exemplar of the intricate emission processes like 
multiple mode changing, nulling and subpulse drifting, in core profiles, which 
have previously been associated with conal only pulsars. To study these 
phenomena in more detail we have carried out sensitive observations of a large 
number of single pulses from this pulsar. We have carefully analysed the single
pulse behaviour to characterise the different phenomena seen in this system. In 
section \ref{sec:obs} we report the observational details; section 
\ref{sec:modnull} details the nulling and mode changing analysis; section 
\ref{sec:pol} presents the polarization behaviour and our estimates of LOS 
geometry; in section \ref{sec:drift} we present the drifting behaviour; section
\ref{sec:disc} presents an in depth discussion of the implications of these 
results on the prevalent understanding of the radio emission process and 
finally we summarize our results in section \ref{sec:sum}.

\section{Observing Details}\label{sec:obs}
\noindent
As mentioned above the pulsar J2006$-$0807 has been observed as part of MSPES,
which recorded polarized single pulses using the Giant Meterwave Radio 
Telescope (GMRT), located near Pune in India \citep{swa91}. The observing 
details of MSPES is reported in \citet{mit16}. The survey observed 123 pulsars 
at two frequencies, 333 and 618 MHz, over a 16 MHz bandwidth, for roughly 2100 
pulses. The pulsar J2006$-$0807 has a period of around 0.58 seconds and a low 
dispersion measure of 32.39~pc~cm$^{-3}$. The MSPES observations were sensitive
to detect single pulses at 333 MHz for this pulsar, while only the average 
profile was measured at 610 MHz. We have carried out further observations at 
339 MHz using the GMRT to characterize the emission properties of PSR 
J2006$-$0807~in more detail. These observations were conducted on 18 November
2017, and recorded around 4800 single pulses. In order to record more sensitive
single pulses we observed using a 33 MHz bandwidth. However, this implied that 
the full polarization signals could not be recorded without significant loss 
due to the higher data rates, and hence we only observed in the total intensity
mode for the newer observations. A number of detailed steps were followed to 
convert the recorded signals to a frequency averaged single pulse sequence. 
This included removing the radio frequency interference from the time sequence 
as well as frequency channels, averaging across the frequency band after 
correcting for the dispersion spread, using the known dispersion measure, and 
finally re-sampling the band averaged data to two dimensional pulse stack with 
the x-axis corresponding to the longitude bins and the y-axis the pulse 
numbers. The details of observational techniques as well as the initial 
analysis is presented in our earlier works \citep[see e.g.][]{bas16,bas17,
bas18a}. The final pulse stack was used for the different analysis described in
the subsequent sections. In addition, we also used the polarization information
from the 333 MHz MSPES observations to estimate the pulsar geometry and 
emission properties as reported in section 4.

\section{Mode Changing and Nulling}\label{sec:modnull}
\subsection{The Emission Modes}\label{subsec:mod}
\begin{figure*}
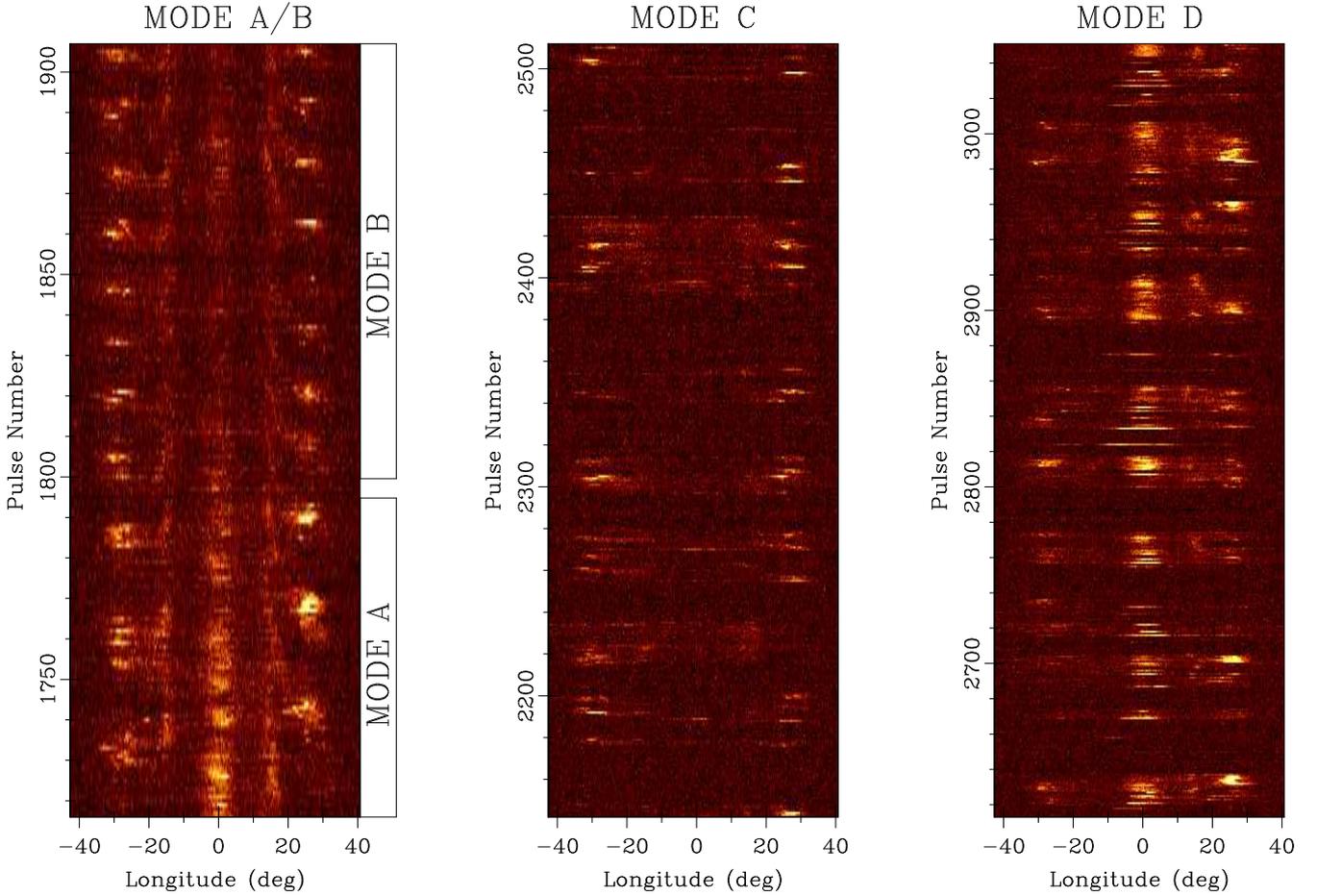

\begin{tabular}{@{}cr@{}cr@{}}
{\mbox{\includegraphics[scale=0.5,angle=0.]{ModeAB_1715-1906.ps}}} &
\hspace{20px}
{\mbox{\includegraphics[scale=0.5,angle=0.]{ModeC_2141-2511.ps}}} &
\hspace{30px}
{\mbox{\includegraphics[scale=0.5,angle=0.]{ModeD_2612-3050.ps}}} \\
\end{tabular}
\caption{The figure represents the different emission modes seen in PSR 
J2006$-$0807. The two modes A and B show the presence of systematic subpulse 
drifting with different periodicities and mode B is usually seen following mode
A. The left window represents a section of the pulse stack ranging from pulse 
number 1716 to 1907, and shows the two modes A and B. Mode A gradually 
transitions into mode B without a clear boundary. Though no clear nulling is 
seen during this sequence, it is seen in the pulses immediately preceding and 
following this region. The middle panel shows the sequence of pulses ranging 
from pulse number 2142 to 2512 and represents mode C. The core emission is 
almost non-existent in this mode which is frequently interspersed with null 
pulses. An example of mode D is shown in the right panel which represents the 
stack from pulse number 2613 to 3051. The emission mode is characterised by the
presence of the strong core component. Similar to previous case, frequent 
nulling is also seen during mode D.}
\label{fig_modesngl}
\end{figure*}

\begin{figure*}
\begin{tabular}{@{}lr@{}}
{\mbox{\includegraphics[scale=0.65,angle=0.]{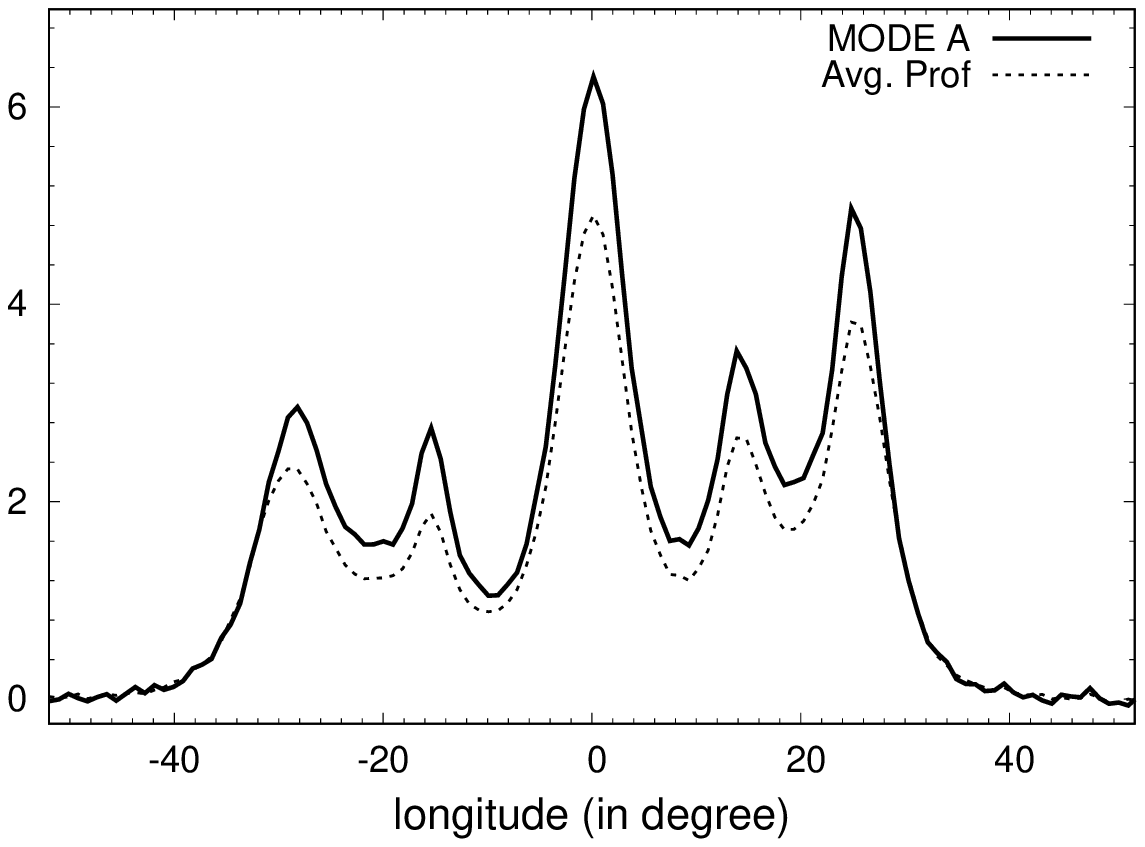}}} &
{\mbox{\includegraphics[scale=0.65,angle=0.]{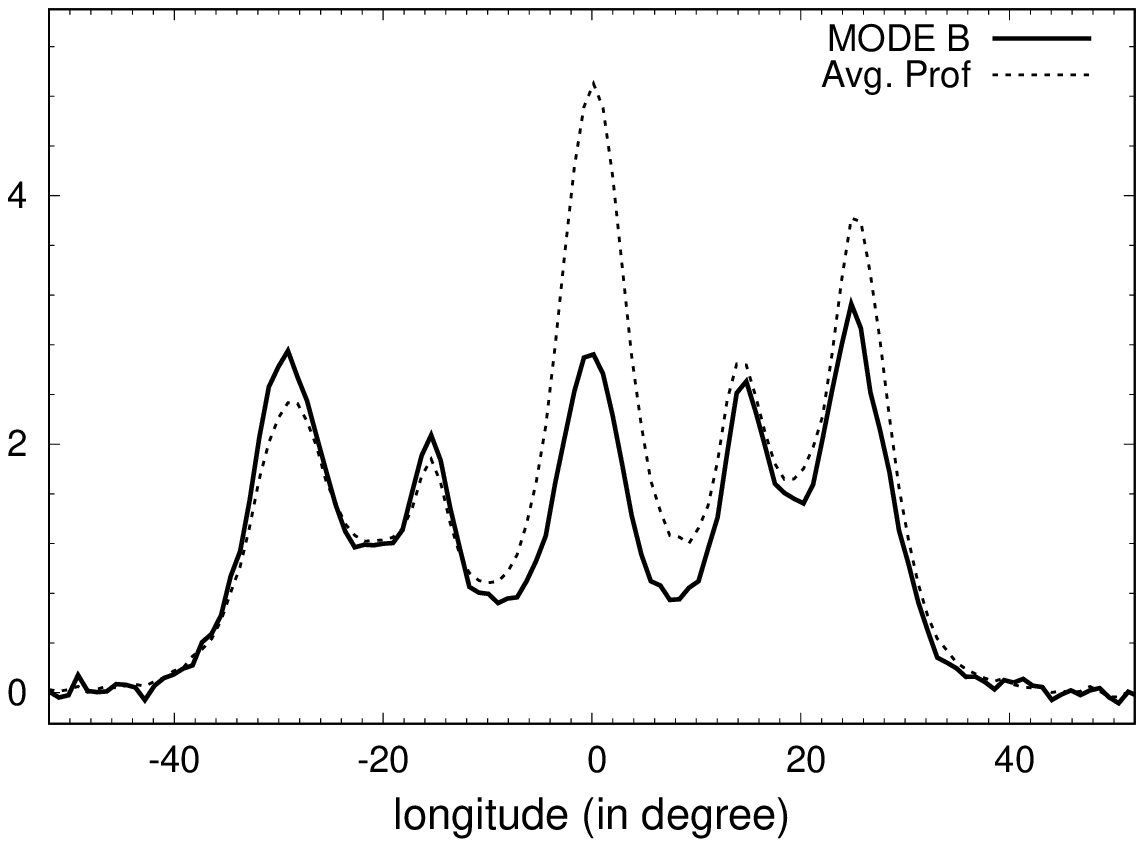}}} \\
{\mbox{\includegraphics[scale=0.65,angle=0.]{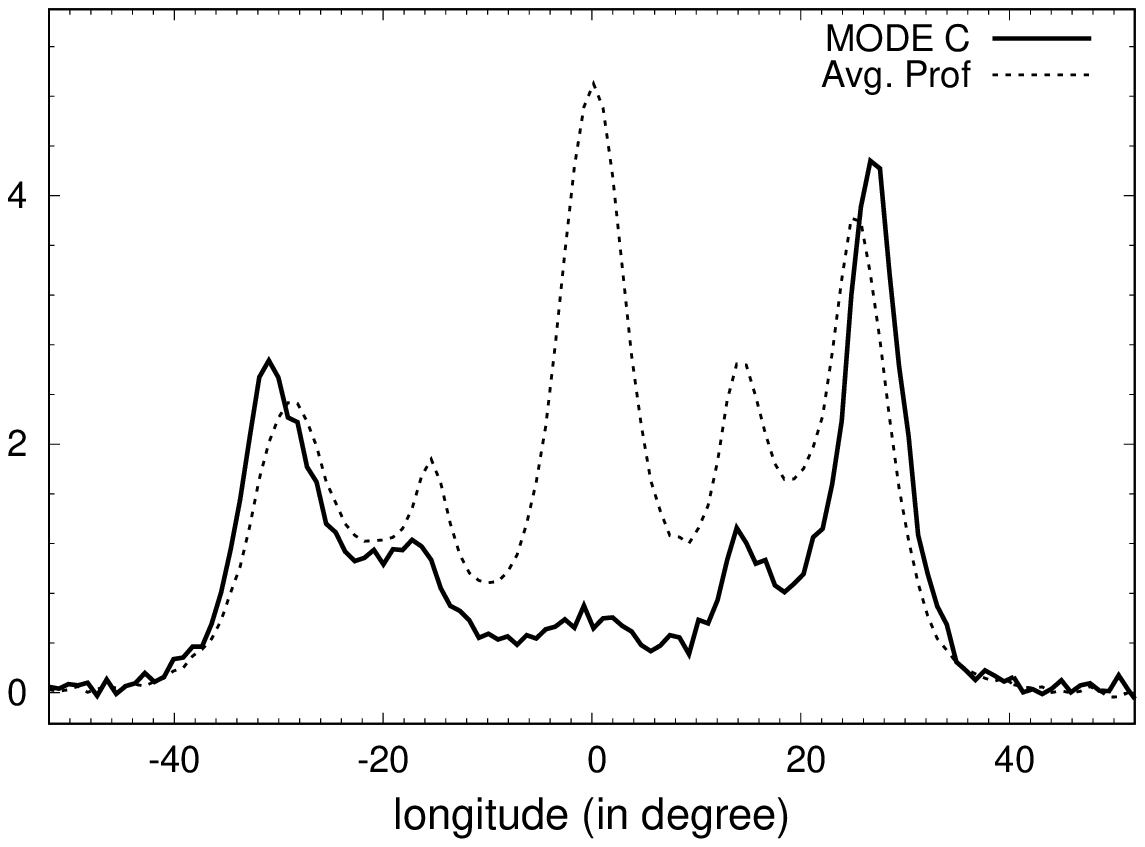}}} &
{\mbox{\includegraphics[scale=0.65,angle=0.]{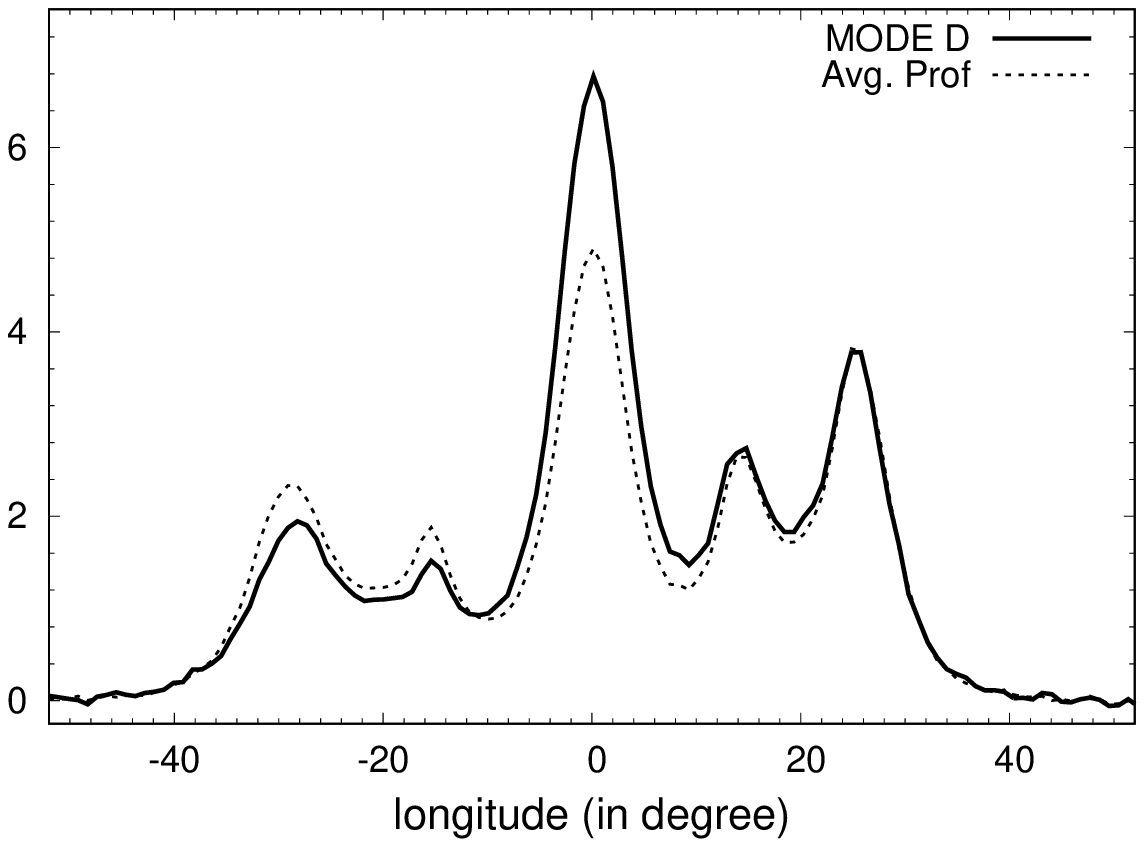}}} \\
\end{tabular}
\caption{The figure shows the profile of each of the four modes present in the 
pulsar J2006$-$0807 along with the average profile for comparison. The top left
panel shows the profile corresponding to mode A, which has a strong core 
emission and the conal components are also prominent. The top right panel 
corresponds to mode B which has a much weaker core component, but the core is 
still comparable to the cones. The bottom left panel shows the profile of mode 
C, where the core is almost non-existent and the inner conal pairs are also 
much weaker. Finally, the mode D profile is shown on the bottom right panel and
is characterised by the strongest core component. The extent of the profiles 
does not change in the different modes, with the outermost extremes coincident 
with the average profile in each case.}
\label{fig_prof}
\end{figure*}

\begin{figure*}
\begin{tabular}{@{}lr@{}}
{\mbox{\includegraphics[scale=0.70,angle=0.]{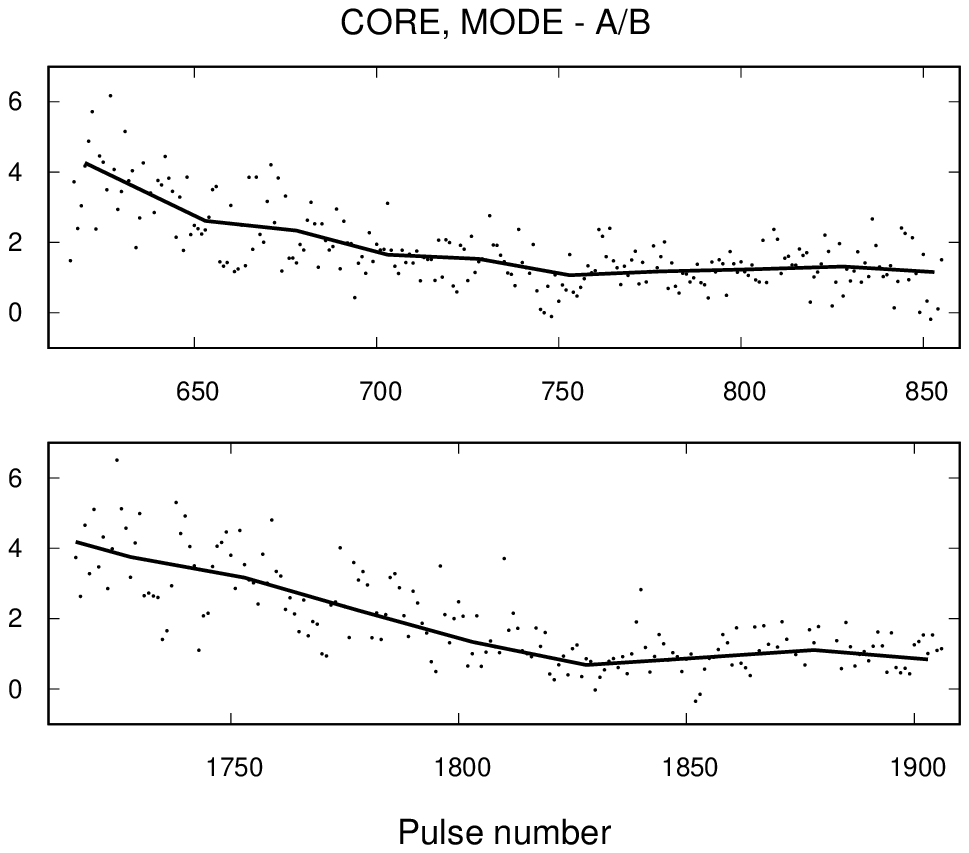}}} &
{\mbox{\includegraphics[scale=0.70,angle=0.]{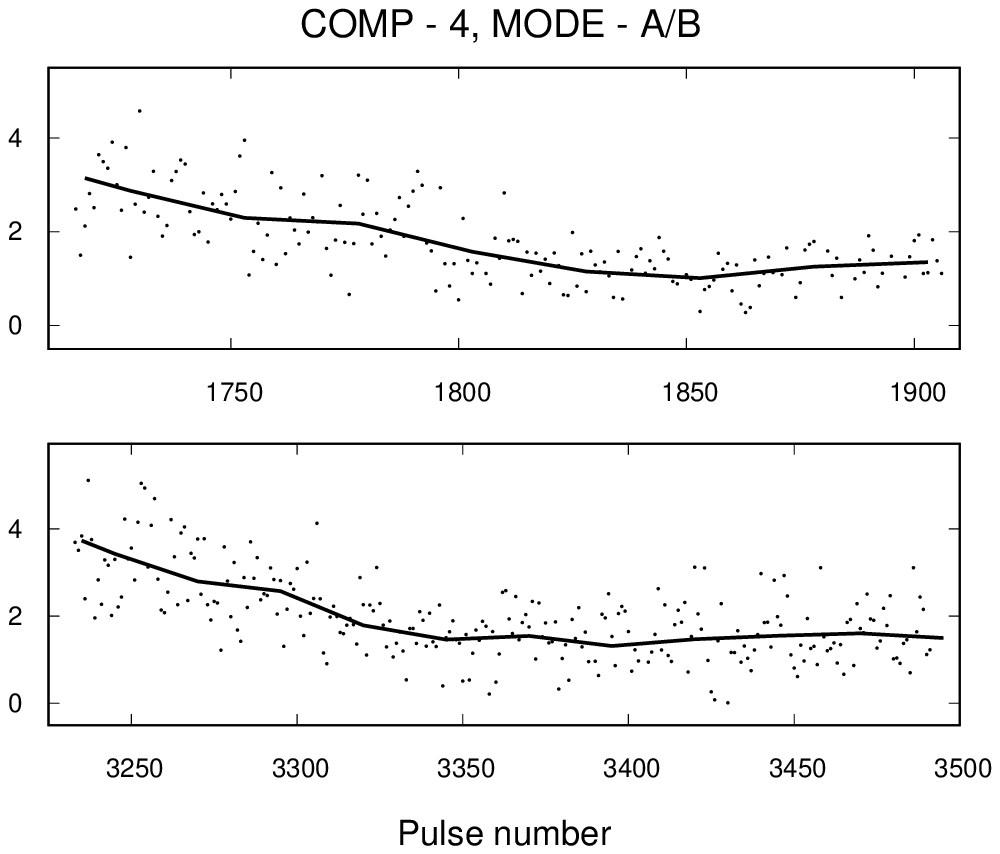}}} \\
\end{tabular}
\caption{The figure shows variation of the component intensities during 
transitions from modes A to mode B. The left panel shows the variation of the 
core energy during two sequences, between pulse 616 and 855 (top window), and 
pulse 1716 and 1907 (bottom window), where the pulsar transitions from mode A 
and B. The average energy in the core window for each pulse is shown as a dot 
while the solid line represents the trend of the energy evolution with time. 
The right panel shows an equivalent plots for the inner conal component 
(4$^{th}$ component), corresponding to two sequences, between pulse 1716 and 
1907 (top window), and pulse 3233 and 3494 (bottom window). The intensities in 
each component window  gradually decreases from the start of mode A and 
stabilizes to a roughly constant level in mode B.}
\label{fig_compmod}
\end{figure*}

\noindent
A wide variety of single pulse behaviour was seen in this pulsar, with frequent
transitions from one emission mode to another. We have carried out a rigorous 
evaluation of the mode changing phenomenon by careful inspection of the single 
pulse sequence. In most of the cases we were able to identify the emission 
modes by visually inspecting them. But there were instances where the modal 
boundaries were overlapping, or the pulsar nulling made it difficult to make 
such distinctions. Short duration sequences ($<$ 10 periods) were not 
classified as distinct modes due to the apparent ambiguity in such 
identifications. We have identified four clear emission modes as shown in 
figure \ref{fig_modesngl}. The most noticeable emission modes were 
characterised by the presence of subpulse drifting with prominent drift bands 
seen in the inner conal pairs, comprising of the second and fourth component, 
as seen in the left panel of the figure. The drifting behaviour showed a 
gradual transition, with the periodicity changing to almost half of the initial
value. However, the boundary of this transition was not sharp and extended over
several pulse periods. We have classified the drifting state into two modes, 
where the initial longer periodicity drifting at the start of the emission was 
termed as mode A and the later shorter periodic state was identified as mode B.
The outer cones, comprising of the first and fifth component, also showed 
similar periodicity during the two modes, but the fluctuations were more phase 
stationary without any clear drift bands. A more detailed analysis of the 
drifting behaviour of modes A and B is presented in section \ref{sec:drift} 
where we have also listed the pulse ranges when the modes are seen in the 
single pulse sequence. In addition to the conal variations the central core 
emission also underwent significant change during the two modes. As shown in 
the average modal profiles in figure \ref{fig_prof} (top left panel), the core 
emission was much stronger during mode A and became comparable to the conal 
emissions in mode B (see the top right panel of the figure \ref{fig_prof}). The
conal emissions were also brighter in mode A compared to mode B. During the 
4800 periods of observations we found three sequences, similar to left panel in
figure \ref{fig_modesngl}, where the two modes coexisted. These sequences 
lasted between 200 and 300 pulses, with mode A typically spanning the first 100
pulses. In figure \ref{fig_compmod} we show variation in the intensities 
of the core and inner conal component at the single pulse level, as the pulsar 
transitioned from mode A to mode B. Both the core and the conal components were
brighter at the start of mode A and the intensity gradually decreased during 
the mode. During mode B a roughly constant intensity with time was seen in the 
different components. In addition there were another eight sequences where the
mode A was seen lasting between 40-100 periods, but instead of changing over to
mode B the pulsar started nulling or moved to a different emission mode 
immediately afterwards. Additionally, another three sequences were found, 
separated by long nulls between pulse number 4050 and 4365, where only mode B 
was seen for durations between 40 and 80 pulses, without the preceding mode A. 
To summarize, during this observation the pulsar lasted for around 15\% of time
in mode A with an average mode duration of 65 periods. The corresponding 
abundance of mode B was around 13\% and the average modal length was roughly 
103 periods. The modal behaviour in the MSPES, 333 MHz observations of roughly 
2100 periods, were very different. It was difficult to ascertain the emission 
modes due to the lower sensitivity of the detections. However, we found one 
long sequence of emission in mode B lasting around 510 periods, preceded by a 
roughly 50 period long mode A. This was found near the beginning of the 
observations. No other clear drifting state was found for the remainder of 
these observations. This also implied that mode B had the possibility of 
lasting for much longer durations than seen in the later observation. 

\begin{table}
\resizebox{\hsize}{!}{
\begin{minipage}{80mm}
\caption{Emission Mode durations in PSR J2006$-$0807}
\centering
\begin{tabular}{ccc}
\hline
 Pulse Range & MODE & Duration \\
  (P) &   & (P) \\
\hline
 ~~12 - ~~33 &  D  & ~22 \\
 ~~67 - ~~83 &  C  & ~17 \\
 ~100 - ~160 &  A  & ~61 \\
 ~211 - ~232 &  C  & ~22 \\
 ~273 - ~472 &  D  & 200 \\
 ~505 - ~528 &  C  & ~24 \\  
 ~569 - ~592 &  A  & ~24 \\  
 ~603 - ~613 &  D  & ~11 \\  
 ~616 - ~856 & A-B & 241 \\  
 ~910 - ~953 &  A  & ~44 \\  
 ~954 - 1376 &  D  & 423 \\  
 1425 - 1455 &  A  & ~31 \\  
 1491 - 1715 &  D  & 225 \\  
 1716 - 1907 & A-B & 192 \\  
 1981 - 2022 &  D  & ~42 \\  
 2031 - 2122 &  A  & ~92 \\  
 2142 - 2512 &  C  & 371 \\  
 2618 - 3229 &  D  & 612 \\  
 3233 - 3494 & A-B & 262 \\  
 3525 - 3592 &  D  & ~68 \\  
 3641 - 3784 & A-B & 144 \\  
 3688 - 3700 &  D  & ~13 \\  
 3795 - 3813 &  C  & ~19 \\  
 3851 - 3885 &  D  & ~35 \\  
 3942 - 3968 &  A  & ~27 \\  
 3972 - 3996 &  D  & ~25 \\  
 4045 - 4057 &  D  & ~13 \\  
 4058 - 4130 &  B  & ~73 \\  
 4137 - 4154 &  C  & ~18 \\  
 4188 - 4240 &  B  & ~53 \\  
 4379 - 4439 &  C  & ~61 \\  
 4519 - 4528 &  D  & ~10 \\  
 4539 - 4554 &  C  & ~16 \\  
 4585 - 4600 &  D  & ~16 \\  
 4614 - 4623 &  C  & ~10 \\  
 4702 - 4798 &  A  & ~97 \\  
\hline
\end{tabular}
\label{tabmodlen}
\end{minipage}
}
\end{table}

In addition to the distinct drifting modes the pulsar also showed the presence 
of two emission states which did not show any drift bands. However, there were 
still periodic modulations associated with these states which are explored in 
the next subsection \ref{subsec:null}. The third mode C was characterised by 
the absence of a clear core emission which was very weak. In figure 
\ref{fig_modesngl}, middle panel, a roughly 370 period duration of the pulse 
sequence in mode C is shown. The emission was frequently disrupted by the 
presence of nulls. The emission was only seen for short durations with the 
minimum being a few periods and the maximum duration of roughly thirty periods.
The average duration of this mode was around 16 periods and the modal abundance
was around 7\%. In figure \ref{fig_prof}, bottom left panel, the average 
profile of mode C is shown. In addition to the lack of a proper core emission, 
the profile also reveals that the inner conal pairs were also significantly 
weaker. Only the outer cones were comparable in intensity with the other modes. 
Finally, the fourth distinct mode present in the pulsar, mode D, was 
characterised by the presence of strong core emission. In figure 
\ref{fig_modesngl}, right panel, a section of the sequence of around 440 pulses
belonging to this mode is shown, which clearly highlights the strong core 
component. Similar to mode C, and unlike modes A and B, this mode also showed 
interspersed null pulses. The mode durations were also similar to mode C, with 
minimum duration of a few periods and maximum of around 35 consecutive pulses.
The average mode duration was 15 periods and the pulsar existed for roughly 
22\% of time in mode D. The average profile of this mode, as shown in 
\ref{fig_prof}, bottom right panel, also reveals that in addition to a strong 
component, the strongest of the three modes, the conal emission was also 
slightly asymmetric. The trailing pair of conal components, corresponding to 
the fourth and fifth components, were more prominent than the leading ones, 
similar to mode A. Table \ref{tabmodlen} lists the pulse sequences during 
each emission mode along with the duration of the modes\footnote{The processed
single pulse data has been made publicly available and can be downloaded from :
ftp://ftpnkn.ncra.tifr.res.in/dmitra/J2006-0807/}. In case there were nulls 
between emission states in the same mode, they have been included in the modal 
duration estimates. However, the nulls on the boundary of two emission modes 
were were not included. In Table \ref{tabmodewid} we have summarized the modal 
behaviour of the pulsar, including the relative abundance, the average duration
of each mode as well as the 50\% width (W$_{50}$) and 5$\sigma$ width 
(W$_{5\sigma}$) of the mode profiles. The profile widths are largely similar 
for the different modes indicating similar emission altitudes.

\begin{table}
\resizebox{\hsize}{!}{
\begin{minipage}{80mm}
\caption{Characterising the emission modes in PSR J2006$-$0807}
\centering
\begin{tabular}{ccccc}
\hline
 MODE & \% & Avg. Duration & W$_{50}$ & W$_{5\sigma}$ \\
  &  & (P) & (\degr) & (\degr) \\
\hline
 A & 15 & ~65 & 60.9$\pm$1.3 & 73.9$\pm$1.3 \\
 B & 13 & 103 & 62.1$\pm$1.3 & 72.4$\pm$1.3 \\
 C & ~7 & ~16 & 64.4$\pm$1.3 & 73.2$\pm$1.3 \\
 D & 22 & ~15 & 62.0$\pm$1.3 & 76.8$\pm$1.3 \\
\hline
\end{tabular}
\label{tabmodewid}
\end{minipage}
}
\end{table}

\subsection{Nulling}\label{subsec:null}
\begin{figure}
\begin{center}
\includegraphics[scale=0.45,angle=-90.,origin=c]{energy_hist.ps}\\ 
\includegraphics[scale=0.66,angle=0.,origin=c]{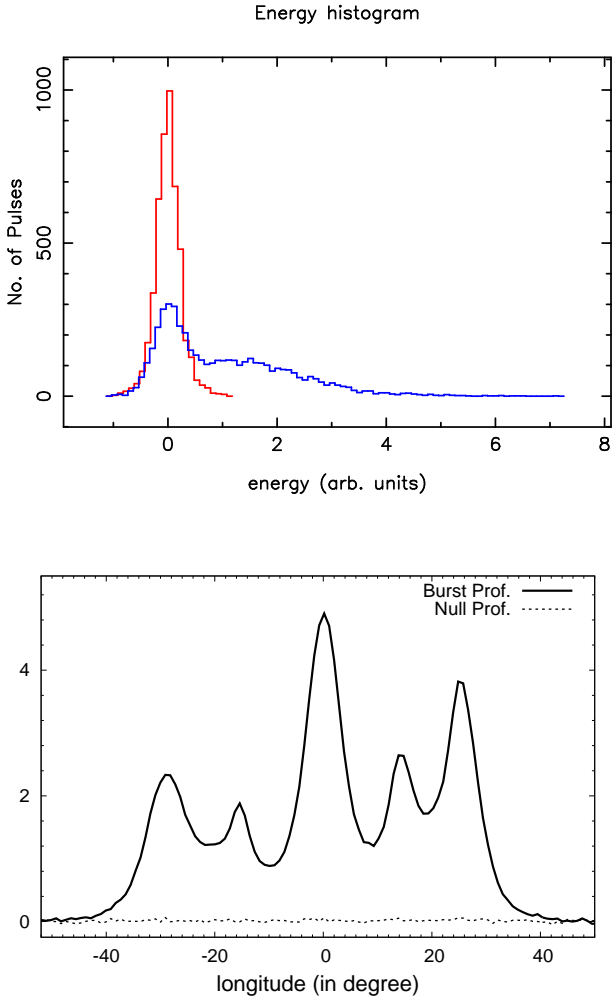}\\
\end{center}
\caption{The figure shows the nulling analysis for the pulsar J2006$-$0807. The
average energy distribution corresponding to the pulse window (blue) as well as
the off-pulse noise (red) is shown on the top window. The on-pulse energies 
show a bimodal distribution with the null pulse distribution coincident with 
the off-pulse. In the bottom panel of the figure we have shown the folded 
profile corresponding to the null and burst pulses. The null profile is noise 
like which indicates that no low level emission was present during nulling.}
\label{fig_nulldist}
\end{figure}

\begin{figure}
\begin{center}
\includegraphics[scale=0.68]{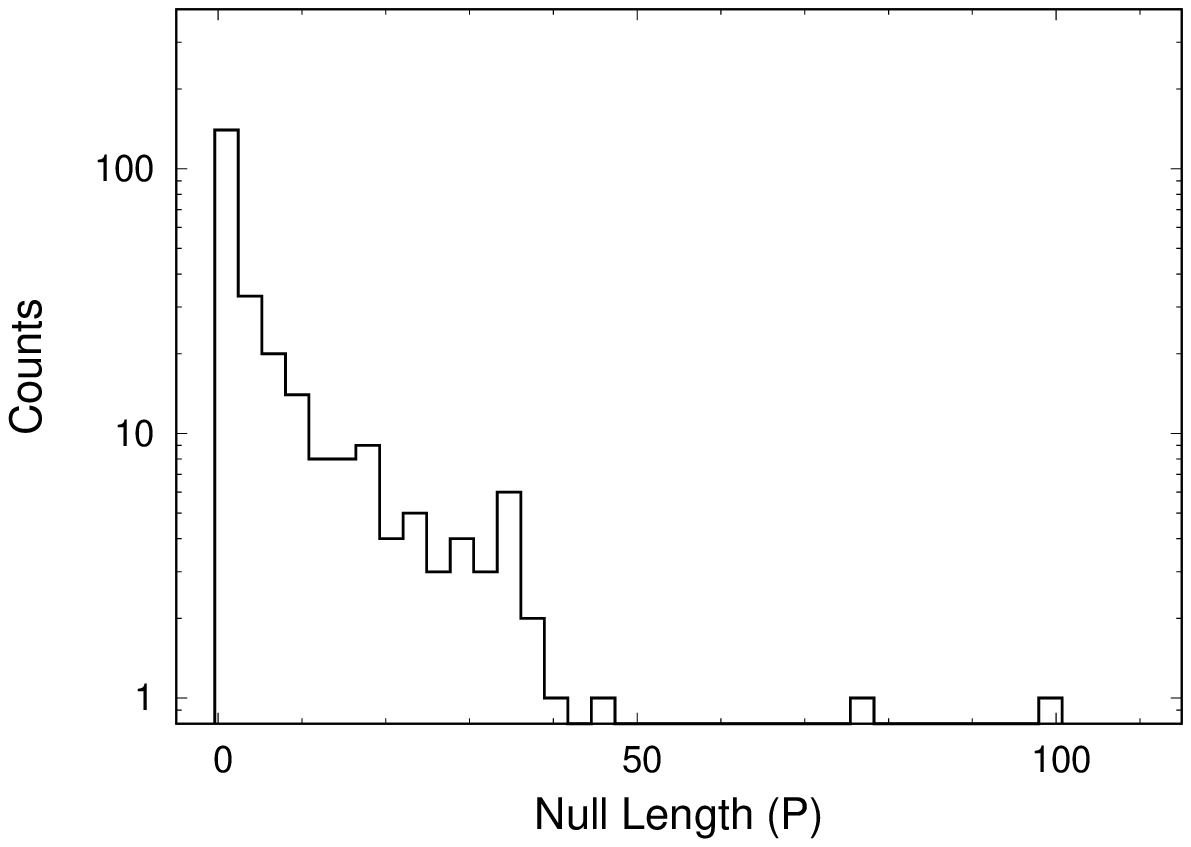}\\ 
\includegraphics[scale=0.68]{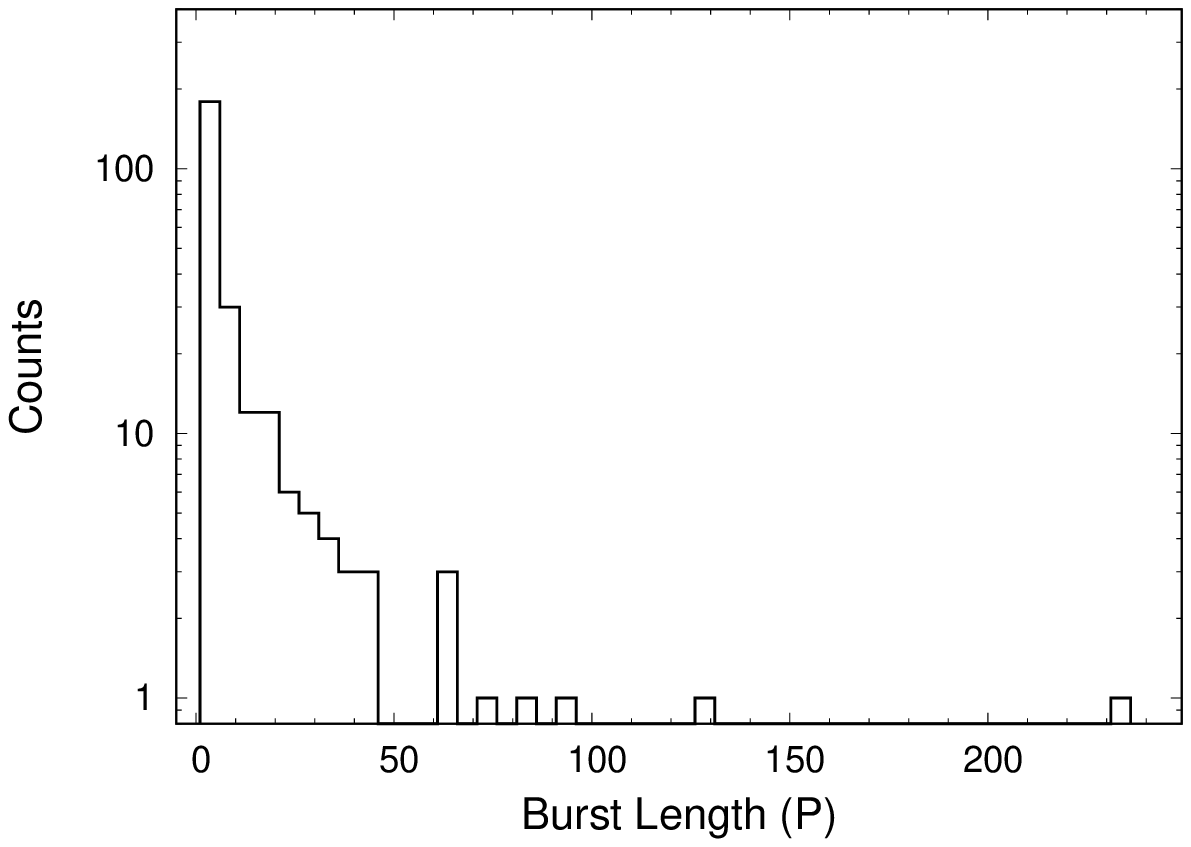}\\
\end{center}
\caption{The figure shows the distributions of the null length (top panel) and
burst length (bottom panel) for the duration of our observations. The 
distributions are dominated by short duration nulls and bursts. The null length
histogram shows evidence of long intervals of nulls lasting 50-100 periods at a
time. The burst length histogram also shows several long intervals ranging from
50 to 200 periods. These correspond to the modes A and B of the pulsar. The 
lower range below 20 periods, on the other hand, are mainly due to modes C and 
D.}
\label{fig_nullen}
\end{figure}

\noindent
The nulling in PSR J2006$-$0807 was reported for the first time in 
\citet{bas17} using the MSPES data. The nulling fractions were reported to be 
15.6$\pm$1.0\% at 333 MHz and 24.2$\pm$1.5\% at 618 MHz, respectively, which 
indicates a wide variation. However, due to the lower sensitivity of these 
observations no further analysis could be conducted. We have used the more 
sensitive 18 November, 2017, observations to carry out a detailed nulling 
analysis of this pulsar employing the methods detailed in \citet{bas17,bas18b}.
The energy distributions for the single pulses and a suitably selected 
off-pulse window were estimated, as shown in figure \ref{fig_nulldist} (top 
window). The on-pulse energies showed a double peaked structure with the 
null pulses showing a peak coincident with the off-pulse distribution. We 
estimated suitable Gaussian approximations for the off-pulse as well as the 
null energy distributions which were used to estimate the nulling fraction. The
individual null and burst pulses were initially identified from the statistical
boundary of the off-pulse distribution. Subsequently, all statistically 
identified nulls were visually inspected to search for low level emission and 
any such cases were reclassified as burst pulses. The nulling fraction was 
36.2$\pm$2.8\% for the 339 MHz observations. This value is higher than the 
previous measurements. As reported in the previous subsection \ref{subsec:mod},
the MSPES observations at 333 MHz contained a long duration of mode B, with 
lower incidence of nulling compared to mode C and D. This likely resulted in 
underestimation of the nulling fraction. In case of the 618 MHz observations, 
the nulling fraction is higher than the 333 MHz value but lower than our latest
measurements. However, the single pulses in this case were too weak to explore 
the modal behaviour. The combined nulling fraction of the three observations is
28.6$\pm$3.4\%. In the bottom window of figure \ref{fig_nulldist} we show the 
profiles formed after averaging the null and burst pulses separately. There is 
no detectable emission seen coinciding with any of the components during 
nulling which justifies our identification scheme for the null and burst 
pulses. The amount by which the radio emission is decreased during nulling is 
given by the factor $\eta$. It is estimated as $\eta$ = $\Sigma 
P(i)$/3$\sigma_N$, here $P(i)$ is the measured intensity along the 
$i^{th}$ bin of the pulse window for the burst profile and $\sigma_N$ is the 
rms in the equivalent window of the null profile. The estimated value is $\eta$
= 2577.5, which is one of the highest reported in the literature \citep[see][]{
viv97,gaj12}. 

The high significance of the single pulse detection during the 18 
November, 2017 observations at 339 MHz enabled us to estimate the durations of
the null and burst events. There were around 260 transitions between nulling 
and bursting during the 4800 periods. In figure \ref{fig_nullen} the 
distributions of the null and burst lengths are shown. The two distributions 
are dominated by short duration nulls and bursts which are consistent with 
previously measured distributions \citep[see][]{bas17}. In addition longer 
duration nulls of 50-100 periods was seen in a few cases. As discussed in the 
previous subsection \ref{subsec:mod}, the longer duration bursts between 50 and
250 periods occurred during modes A and B. The other two modes C and D were 
frequently interrupted by short nulls. The average null length was 8.3 periods,
while the average burst length was of 9.8 period duration.

\begin{figure*}
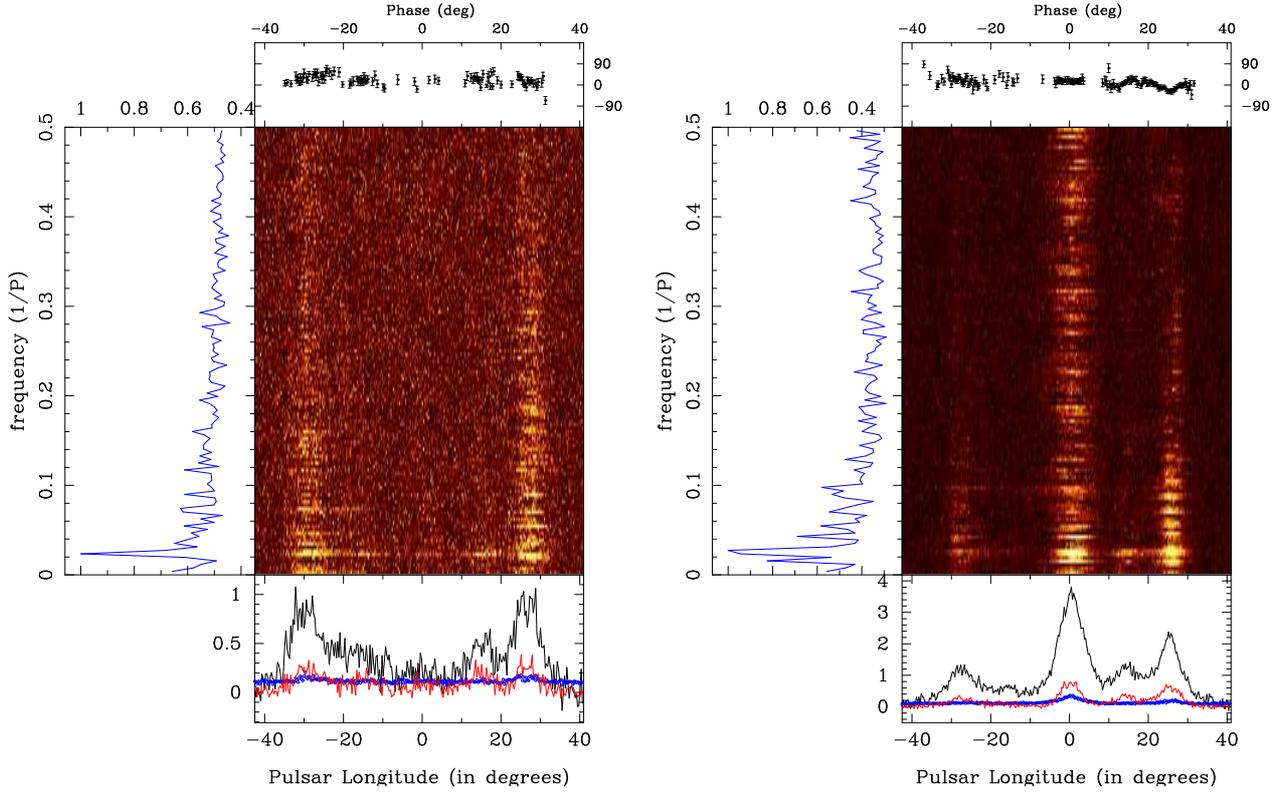

\begin{tabular}{@{}lr@{}}
{\mbox{\includegraphics[scale=0.42,angle=0.]{ModeC_permod.ps}}} &
{\mbox{\includegraphics[scale=0.42,angle=0.]{ModeD_permod.ps}}} \\
\end{tabular}
\caption{The figure shows the Longitude Resolved fluctuation spectra (LRFS) for
two separate pulse sequences of the pulsar J2006$-$0807. The left panel 
corresponds to the sequence between pulse number 2142 and 2397, which belongs 
to mode C. The right panel shows an equivalent plot between pulse number 2618 
and 2873, when the pulsar is in mode D. The central window shows the 
variation of the fluctuation spectra across the different longitudes. The 
bottom window shows the average profile from the pulses used for the spectra 
(black line). The bottom window also shows the variation of the peak amplitude
across the pulse window (red line) along with the 3$\sigma$ baseline level of
the LRFS along each longitude (blue dotted line). The left window shows the 
average LRFS across all longitudes and clearly indicates the presence of low 
frequency periodicity in both modes. The top window plots the phases across 
each longitude corresponding to all significant peak amplitudes greater than 
the baseline level. The phases show very little variations across the pulse 
window.} 
\label{fig_ampmod}
\end{figure*}

\begin{figure*}
\begin{tabular}{@{}lr@{}}
{\mbox{\includegraphics[scale=0.65,angle=0.]{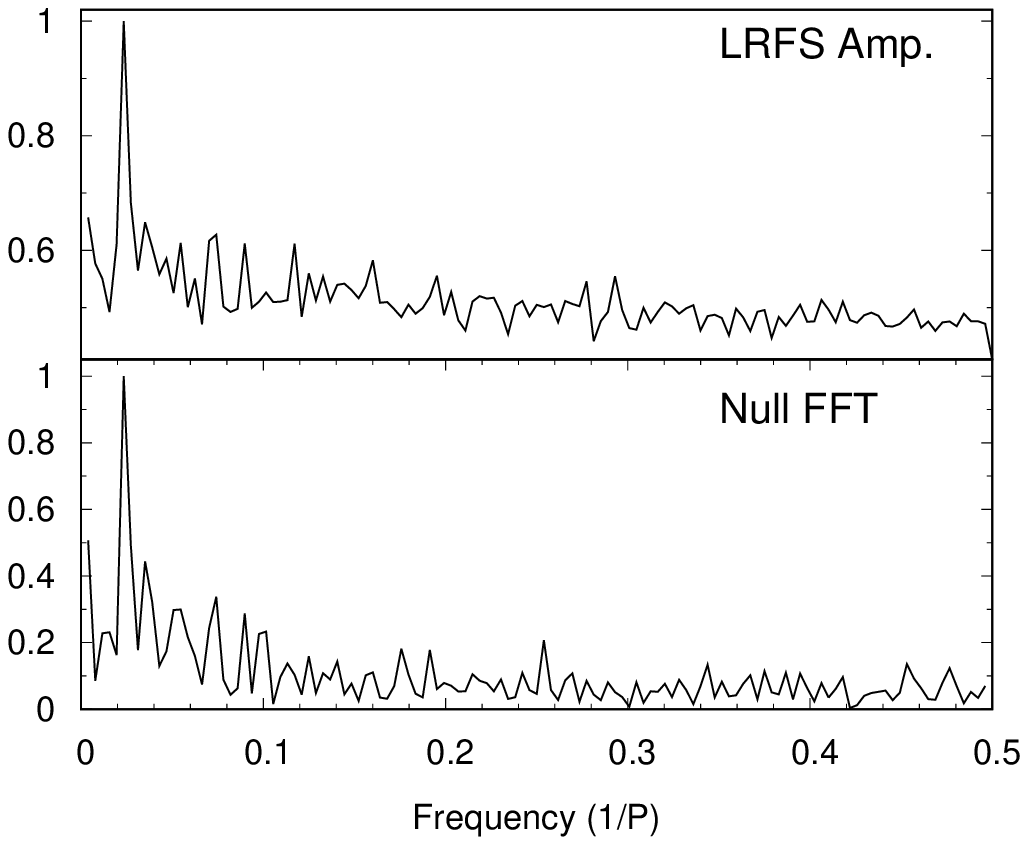}}} &
{\mbox{\includegraphics[scale=0.65,angle=0.]{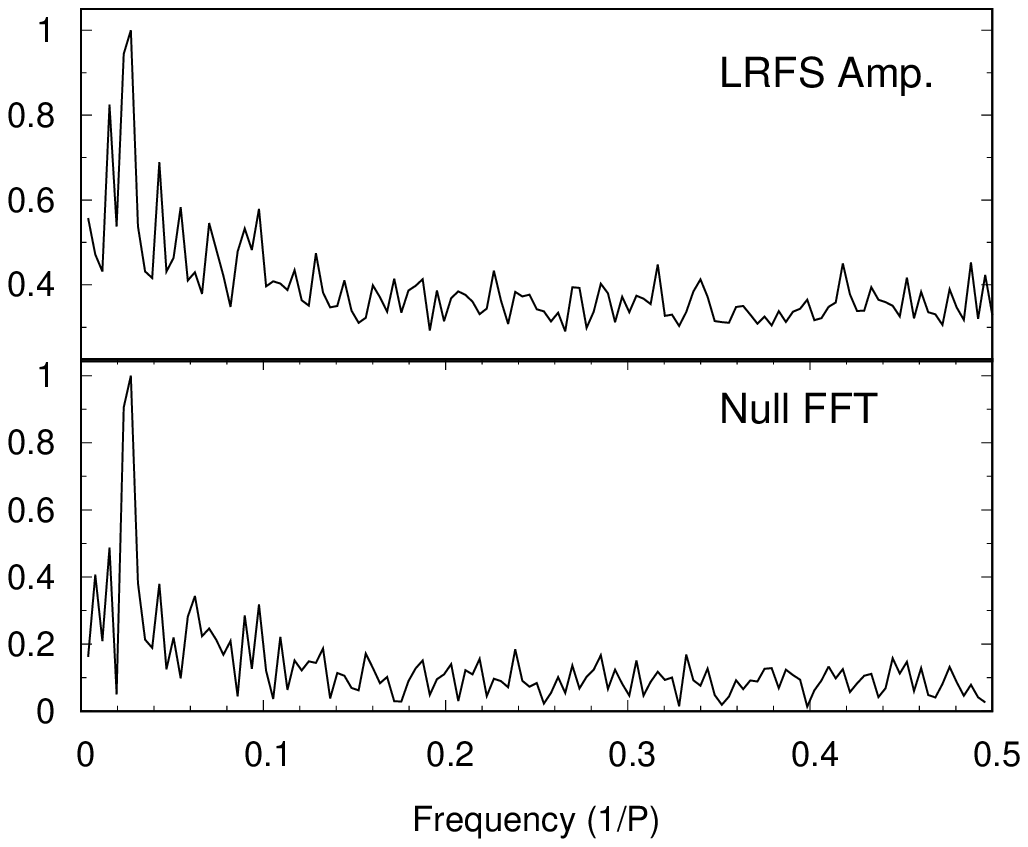}}} \\
\end{tabular}
\caption{The figure shows a comparison between the longitude averaged amplitude
of the Longitude Resolved Fluctuation spectrum (LRFS) of a pulse sequence and 
the spectra of the Fourier transform of the 0/1 series, with 0 corresponding to
any null pulse and 1 corresponding to the burst pulse, for the same sequence. 
The left panel covers the pulse range 2142 and 2397, and belongs to mode C of 
PSR J2006$-$0807. The right panel is constructed for the pulse sequence between 
2618 and 2873 periods and belongs to mode D. The low frequency peak is seen to 
be identical in both the spectra for each mode, which suggests that the low 
frequency structure in the LRFS originates due to nulling.}
\label{fig_nulfft}
\end{figure*}

In addition to the subpulse drifting seen in modes A and B, the pulsar also 
showed a low frequency feature in the average fluctuation spectra 
\citep{bas16}. However, due to the presence of a large number of modes with 
different periodic features, the low frequency feature could not be resolved in
the earlier work. We have now carried out a detailed study of the individual 
modes to investigate the origin of this periodicity in more detail. The 
fluctuation spectra corresponding to mode A and B has been investigated to 
evaluate the drifting behaviour. In case of these two modes no low frequency 
feature was seen and the only periodicity was due to the drifting, which is 
discussed in section \ref{sec:drift}. However, in the case of mode C and D the
low frequency feature was seen in the fluctuation spectra. In certain sections
of these modes clear low frequency peaks could be clearly seen. In figure 
\ref{fig_ampmod} we show two examples of the Longitude Resolved Fluctuation 
Spectra \citep[LRFS,][]{bac70,bac73} corresponding to mode C (left panel), 
between pulse number 2142 and 2397, and mode D (right panel), between pulse 
number 2618 and 2873. The presence of low frequency feature is clearly seen. 
The modulations responsible for these feature are not due to drifting but are 
periodic amplitude modulations. This is clearly demonstrated by the phase 
variations, shown in the top window of each plot, which are relatively flat and
featureless across the entire window. In a number of pulsars the low frequency 
features have now been shown to be a result of periodic nulling \citep{bas17}. 
To investigate this possibility we carried out an analysis similar to 
\citet{bas17}, where a sequence of `0' and `1' were generated corresponding to 
the null and burst pulses respectively, and Fast Fourier Transforms (FFT) of 
these sequences were carried out. In figure \ref{fig_nulfft} we show the 
results of this exercise for the two sequences belonging to mode C and D, 
described earlier, and compare the low frequency features with the longitude 
average LRFS for these sequences. The figure shows that these two features are 
identical, which suggests, that during certain instances the pulsar exhibit 
periodically varying nulling states. The periodic nulling is primarily seen 
during modes C and D, and is absent during the other modes two modes. Due to 
the frequent mode transitions and presence of long duration nulls, the low 
frequency feature becomes diffuse and indistinguishable from the boundary in 
the time average fluctuation spectra. The periodicity of the feature, when 
clearly seen, is $P_N$ = 41.3$\pm$4.2 $P$. The presence of low frequency 
features associated with nulling, and the presence of drifting in the same 
system, has now been detected in eight pulsars, including J2006$-$0807 
\citep{bas17,bas18b}. However, this is the first case where we have been able 
to show the periodic nulling clearly in the presence of core emission, where 
the core and cone nulls simultaneously. 

\begin{table}
\resizebox{\hsize}{!}{
\begin{minipage}{80mm}
\caption{Summary of Nulling in PSR J2006$-$0807}
\centering
\begin{tabular}{cccccc}
\hline
 NF & $\eta$ & $N_T$ & $\langle BL\rangle$ & $\langle NL\rangle$ & $P_N$ \\
 (\%) &   &   & ($P$) & ($P$) & ($P$) \\
\hline
   &   &   &   &   &   \\
 28.6$\pm$3.4 & 2577.5 & 262 & 8.3 & 9.8 & 41.3$\pm$4.2 \\
   &   &   &   &   &   \\
\hline
\end{tabular}
\label{tabnull}
\end{minipage}
}
\end{table}

In Table \ref{tabnull} we have summarized the nulling behaviour in this pulsar.
The Table lists the total nulling fraction (NF) over all the observations, the
reduction in intensity during nulling, specified by $\eta$, the number of 
transitions between null and burst sequences ($N_T$), the average null 
($\langle NL\rangle$) and burst lengths ($\langle BL\rangle$), and the 
periodicity ($P_N$) of the nulling pattern during certain sections of pulse 
sequence.

\section{Polarization Behaviour : Emission Geometry}\label{sec:pol}
\begin{figure*}
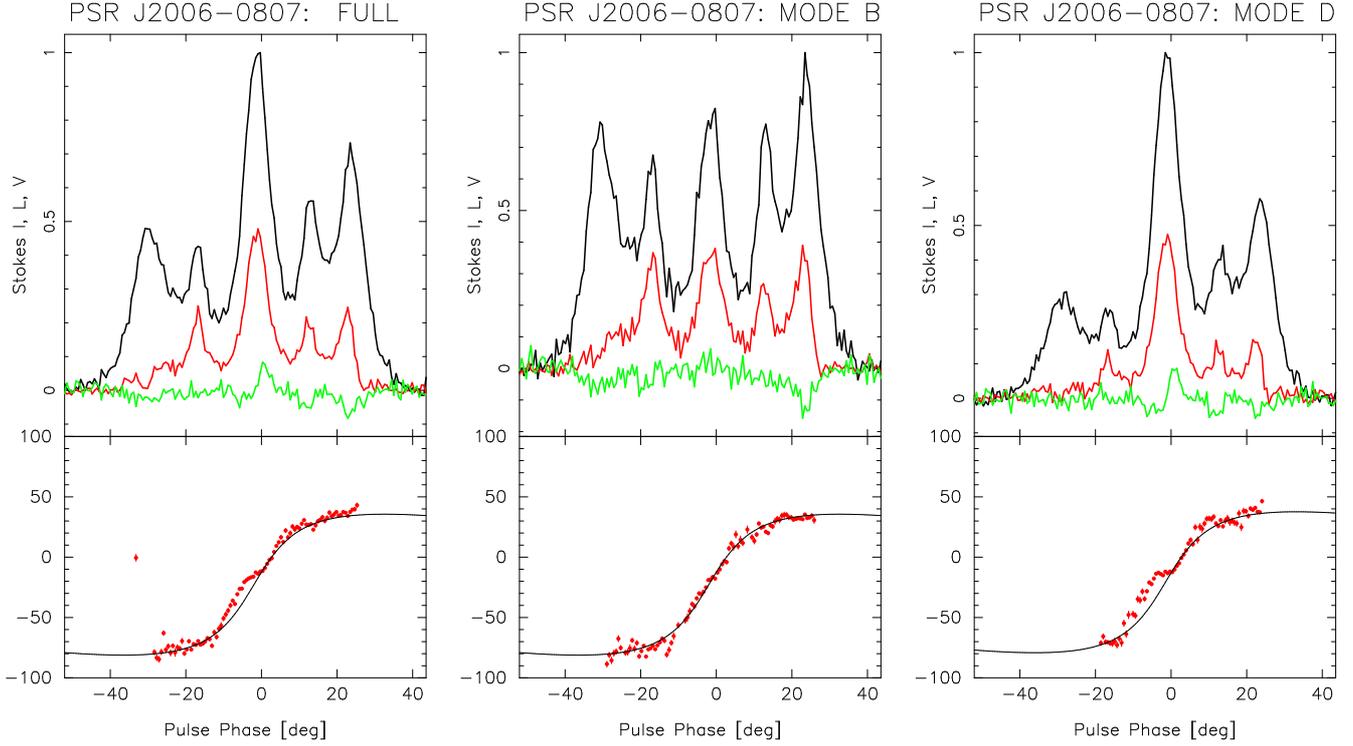

\begin{tabular}{@{}cr@{}cr@{}}
{\mbox{\includegraphics[scale=0.4,angle=0.]{polfull.ps}}} &
{\mbox{\includegraphics[scale=0.4,angle=0.]{polmodeB.ps}}} &
\hspace{10px}
{\mbox{\includegraphics[scale=0.4,angle=0.]{polmodeD.ps}}} \\
\end{tabular}
\caption{The figure shows the average polarization properties of the pulsar
J2006$-$0807 using the MSPES, 333 MHz observations \citep{mit16}. Due to the 
lower sensitivity and shorter duration of these observations we could only
obtain sufficiently sensitive studies of two emission modes, mode B and mode D. 
The left panel corresponds to the average over the full observing session, the
middle panel shows mode B and right panel represents the behaviour of mode D.
The top window of each plot shows the average profile (black line) along with 
the linear polarization (L, red line) and the circular polarization (V, green
line). The bottom window in each figure shows the polarization position angle 
(PPA). The figure also shows the rotating vector model (RVM) fits to the PPA, 
where the RVM fits have been estimated from PPA of mode B.}
\label{fig_modepol}
\end{figure*}

\begin{table*}
\caption{The polarization properties of PSR J2006$-$0807}
\centering
\begin{tabular}{cccccccccc}
\hline
   & Type & \multicolumn{2}{c}{Longitude Range} & \multicolumn{2}{c}{AVG. PROFILE} & \multicolumn{2}{c}{MODE B} & \multicolumn{2}{c}{MODE D} \\
   &  & \multicolumn{2}{c}{} & $\%(L/I)$ & $\%(V/I)$ & $\%(L/I)$ & $\%(V/I)$ & $\%(L/I)$ & $\%(V/I)$ \\
\hline
   &   &   &   &   &   &   &   &   &   \\
 Full Window & --- & -40.3\degr & 34.2\degr & 29.86$\pm$0.02 & -2.93$\pm$0.03 & 30.29$\pm$0.03 & -6.06$\pm$0.05 & 28.23$\pm$0.03 & -1.79$\pm$0.04 \\
   &   &   &   &   &   &   &   &   &   \\
 Component I & Outer Cone & -40.3\degr & -22.0\degr & 12.20$\pm$0.09 & -4.7$\pm$0.1 & 13.8$\pm$0.1 & -7.0$\pm$0.2 & 9.3$\pm$0.2 & -2.8$\pm$0.3 \\
   &   &   &   &   &   &   &   &   &   \\
 Component II & Inner Cone & -21.4\degr & -10.4\degr & 41.7$\pm$0.2 & -1.7$\pm$0.2 & 47.0$\pm$0.2 & -5.7$\pm$0.3 & 33.8$\pm$0.3 & 0.7$\pm$0.5 \\
   &   &   &   &   &   &   &   &   &   \\
 Component III & Core & -9.8\degr & 7.3\degr & 40.91$\pm$0.06 & 1.23$\pm$0.08 & 42.2$\pm$0.1 & -0.4$\pm$0.2 & 39.51$\pm$0.07 & 0.6$\pm$0.1 \\
   &   &   &   &   &   &   &   &   &   \\
 Component IV & Inner Cone & 7.9\degr & 17.1\degr & 31.1$\pm$0.1 & -5.7$\pm$0.2 & 30.5$\pm$0.2 & -6.6$\pm$0.3 & 27.7$\pm$0.2 & -5.1$\pm$0.3 \\
   &   &   &   &   &   &   &   &   &   \\
 Component V & Outer Cone & 17.7\degr & 34.2\degr & 20.92$\pm$0.08 & -5.6$\pm$0.1 & 25.0$\pm$0.1 & -10.4$\pm$0.2 & 17.7$\pm$0.1 & -4.5$\pm$0.2 \\
   &   &   &   &   &   &   &   &   &   \\
\hline
\end{tabular}
\label{tab_pol}
\end{table*}

\noindent
In our latest observations at 339 MHz we were unable to measure the polarized 
single pulses. Hence, we could not explore the polarization properties of the 
different emission modes in great detail. However, we have used the 333 MHz, 
MSPES observations to estimate the average polarization behaviour. As mentioned
in section \ref{sec:obs} the MSPES was not sensitive enough for polarized 
single pulses and we could only explore the average polarization behaviour. 
The survey recorded all the four stokes parameters (I, Q, U, V) for the 
pulsar. In figure \ref{fig_modepol} we show the average polarization behaviour
corresponding to the average profile (left window), the mode B (middle window)
and mode D (right window). We were unable to detect sufficient number of single
pulses in the modes A and C to form stable average profiles. The figure shows
the average and modal profiles as well as the average linear (L = 
(Q$^2$+U$^2$)$^{0.5}$) and circular (V) polarization measurements. Table 
\ref{tab_pol} shows the percentage of linear and circular polarization for the 
average profile as well as the two modes. We have calculated the percentage 
polarization, as a ratio between the linear and circular polarization with
the total intensity, for the entire pulse window as well as the five 
individual components. 
The ratios were estimated as 
$\langle$L($\psi$)$\rangle$/$\langle$I($\psi$)$\rangle$ and 
$\langle$V($\psi$)$\rangle$/$\langle$I($\psi$)$\rangle$, respectively, where 
$\psi$ represents any specific longitude. 
The average linear polarization in all three cases were around 30\%. However, 
this varied considerably across the different components. The core emission 
showed around 40\% linear polarization in both the emission modes. The sign 
changing circular polarization was also seen for this component which is 
consistent with the core behaviour \citep{ran90}. The polarization was 
significantly lower in the first component, with mode B showing around 14\% 
linear polarization and mode D around 9\%. The inner cones, comprising of the 
second and fourth components, had significantly higher polarization, similar to
the core, in both modes. The fifth component, on the other hand, had slightly 
lower polarization of around 25\% in mode B and 18\% in mode D, which was still
higher than the leading component. 

\begin{table*}
\caption{Estimating the emission geometry of PSR J2006$-$0807}
\centering
\begin{tabular}{cccccccccc}
\hline
   & $W_{50}^{core}$ & $W_s^{in}$ & $W_s^{out}$ & $\alpha$ & $\beta$ & $\rho_{in}$ & $\beta$/$\rho_{in}$ & $\rho_{out}$ & $\beta$/$\rho_{out}$\\
   & (\degr) & (\degr) & (\degr) & (\degr) & (\degr) & (\degr) &  & (\degr) & \\
\hline
   &  &  &  &  &  &  &  \\
 Avg. Profile & 8.2$\pm$0.9 & 29.7$\pm$0.9 & 54.0$\pm$0.9 & 22.4$\pm$5.2 & 4.7$\pm$1.0 & 7.8$\pm$0.2 & 0.60 & 12.1$\pm$0.6 & 0.39 \\
   &  &  &  &  &  &  &  &  &  \\
 Mode A & 7.9$\pm$0.9 & 29.3$\pm$0.9 & 53.1$\pm$0.9 & 23.5$\pm$5.6 & 4.9$\pm$1.1 & 8.0$\pm$0.3 & 0.61 & 12.5$\pm$0.6 & 0.39 \\
   &  &  &  &  &  &  &  &  &  \\
 Mode B & 8.2$\pm$0.9 & 30.2$\pm$0.9 & 54.0$\pm$0.9 & 22.5$\pm$5.2 & 4.7$\pm$1.0 & 7.9$\pm$0.3 & 0.59 & 12.2$\pm$0.6 & 0.39 \\
   &  &  &  &  &  &  &  &  &  \\
 Mode C & --- & 31.1$\pm$0.9 & 58.0$\pm$0.9 & --- & --- & 8.0$\pm$0.3 & --- & 12.9$\pm$0.7 & --- \\
   &  &  &  &  &  &  &  &  &  \\
 Mode D & 8.2$\pm$0.9 & 29.7$\pm$0.9 & 53.5$\pm$0.9 & 22.4$\pm$5.2 & 4.7$\pm$1.0 & 7.8$\pm$0.2 & 0.60 & 12.0$\pm$0.6 & 0.39 \\
   &  &  &  &  &  &  &  &  &  \\
\hline
\end{tabular}
\label{tab_geom}
\end{table*}

In addition, we have also estimated the polarization position angle (PPA) which
are shown in the bottom panel of figure \ref{fig_modepol}, for all three cases.
The PPA usually shows a S-shaped curve, as seen in the figure, and is 
interpreted using the rotating vector model \citep[RVM,][]{rad69}, where the 
line of sight traverses diverging magnetic field lines. The PPA also shows the
presence of two orthogonal polarization modes \citep{gil95,mit09}, which are
offset by 90\degr~in phase, and are believed to be associated with the X and O
modes of the emitting plasma waves \citep{mel14,mit17b}. We were not able to 
explore the orthogonal modes from the single pulses but the average PPA showed
signatures of the orthogonal modes. Mode B showed a typical S-shaped curve 
which closely followed the RVM. However, mode D showed significant departure 
from the RVM nature suggesting the presence of orthogonal modes. The deviation 
is seen primarily in the core component which is more prominent in this mode. 
The above behaviour suggests interesting associations between the emission 
modes and orthogonal polarization moding in this pulsar. Similar behaviour 
associated with the core component has also been reported during the different 
modes of PSR B1237+25 \citep{smi13}. This has interesting implications on the
pulsar magnetospheric conditions during mode changing phenomenon and requires 
more sensitive studies of polarized single pulses from PSR J2006$-$0807.

The measurement of the profile widths, as well as the presence of average PPA 
estimates, make it possible to investigate the emission geometry. The geometry 
is usually characterised by two quantities, $\alpha$, which represents the 
angle between the rotation and magnetic axis of the pulsar, and $\beta$, the 
angle between the line of sight and rotation axis during their closest 
approach. It has been shown that in pulsars with a central core component, 
$\alpha$ is related to the core width, sin$\alpha$ = W$_{b}$P$^{-0.5}/W^c$, 
where $W^c$ is the core width measured at 50\% level of peak intensity, and 
W$_{b}$ corresponds to the lower boundary of the distribution of the core 
widths with period \citep{ran90}. The core boundary width have recently 
been measured to be W$_{b}$ = 2.39$\pm$0.26\degr, at 333 MHz, by \citet{skr18}.
The angles $\alpha$ and $\beta$ can also be estimated from the RVM fits to the 
PPA. However, these fits yield values which are highly correlated and are not 
effective in estimating the emission geometry \citep{eve01,mit04}. However, the
steepest gradient (R$_{PPA}$) of the PPA is related to the two angles, 
R$_{PPA}$ = sin$\alpha$/sin$\beta$, and provides a more reliable estimate. 
Once, the geometrical angles are known, the opening angle ($\rho$) of the 
emission beam can also be estimated from the separation between the inner and 
outer conal component pairs using principles of spherical geometry 
\citep{gil84}, cos$\rho$ = cos$\beta$ - 
2sin$\alpha$~sin($\alpha$+$\beta$)~sin$^2$W$_{sep}$/4.\\

\begin{figure*}
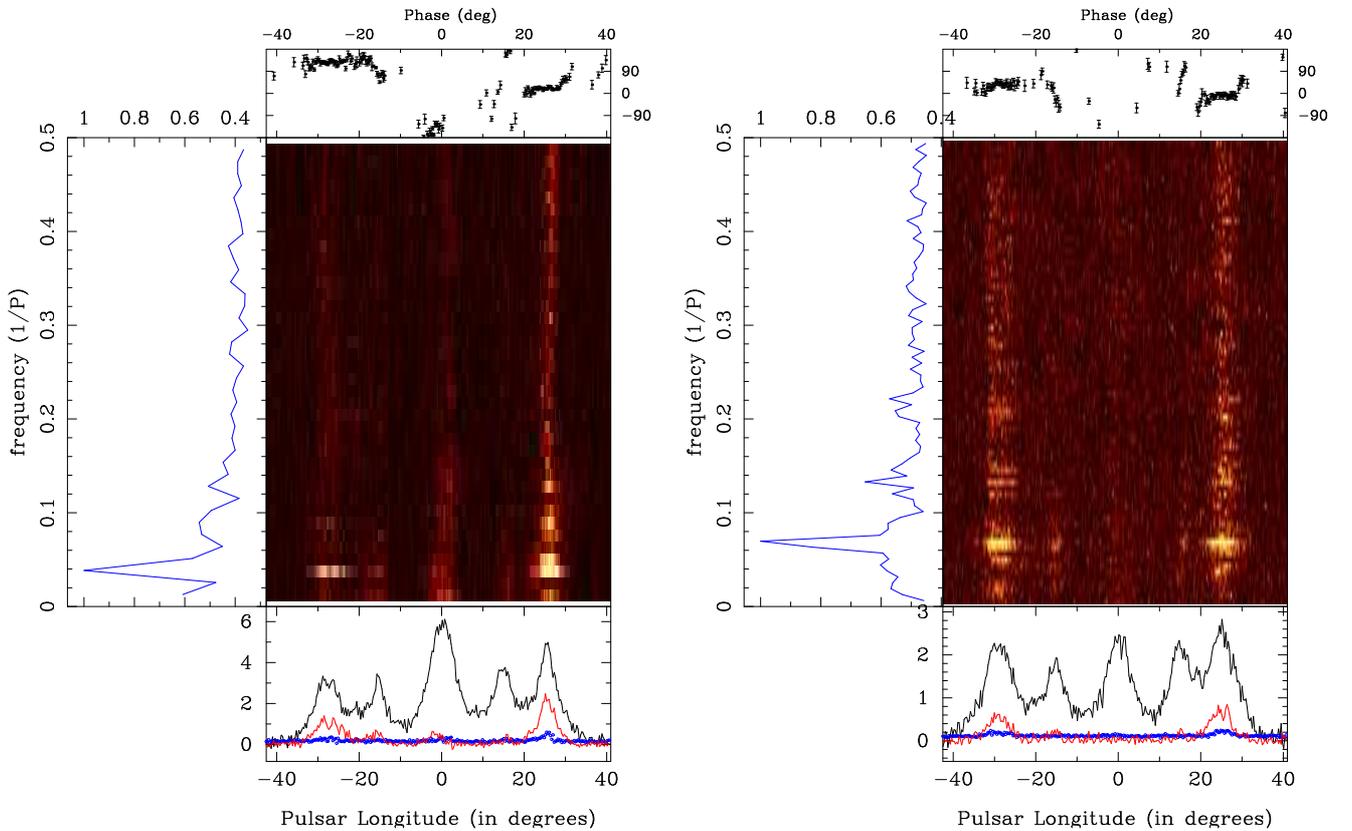

\begin{tabular}{@{}lr@{}}
{\mbox{\includegraphics[scale=0.44,angle=0.]{ModeA_lrfs.ps}}} &
{\mbox{\includegraphics[scale=0.44,angle=0.]{ModeB_lrfs.ps}}} \\
\end{tabular}
\caption{The figure shows the Longitude Resolved fluctuation spectra (LRFS) for
two separate pulse sequences of the pulsar J2006$-$0807. The left panel
corresponds to the sequence between pulse number 1716 and 1793, which belongs
to mode A. The right panel shows an equivalent plot between pulse number 697
and 855, when the pulsar is in mode B. The subpulse drifting is seen in both 
modes with different periodicities. The phase variations are also shown in the 
top window, and shows clear evolution between the inner and outer cones.  
The lower window shows the average profile formed from the pulses used for the
LRFS estimates in each case (black lines). The bottom window also shows the 
variation of the amplitude of the peak frequency across the pulse window (red 
line) as well as 3$\sigma$ level of the baseline of the LRFS for each pulse 
longitude (blue dotted line).}
\label{fig_lrfsdrift}
\end{figure*}

In Table \ref{tab_geom} we have estimated the emission geometry for the average
profile as well as the different emission modes. The Table shows the estimated
core widths at 50\% level of peak intensity, the separation between the peaks 
of the inner ($W_s^{in}$) and outer conal ($W_s^{out}$) pairs, as well as the 
estimated $\alpha$ from the core width. In order to estimate $\beta$ we 
calculated R$_{PPA}$ = 4.6$\pm$0.2~\degr/\degr, using the PPA of mode B which 
showed least deviations from the RVM. The $\beta$ for all the modes are shown 
in the Table along with the $\rho$ for the inner and outer cones. The RVM 
fits shown in figure \ref{fig_modepol} were estimated from the geometry 
reported in Table \ref{tab_geom}. In case of mode C the core component was not
prominent and hence the geometry could not be estimated. However, we calculated
the opening angles for the conal separation using average geometry. The pulsar 
geometry gives similar values for the different emission modes.
Our estimates of $\alpha$ were different from the previous measurement of 
\citet{ran93b} who reported a value of $\alpha$=13\degr. This primarily results
from the estimate of the core width, which was 14\degr~in the previous work, 
and significantly larger than our measurements. We examined the high frequency 
($>$ 1 GHz) profiles available for this pulsar, from the archival observations,
and it was found that the core was not clearly separated from the conal 
emission at higher frequencies. This can possibly lead to a larger estimate. It
is also possible that the pulsar widths have underwent an absorption at lower 
frequencies leading to narrower measurements \citep{ran83}. This will require 
more detailed observations particularly at frequencies above 1 GHz.

\section{Subpulse Drifting}\label{sec:drift}

\begin{figure*}
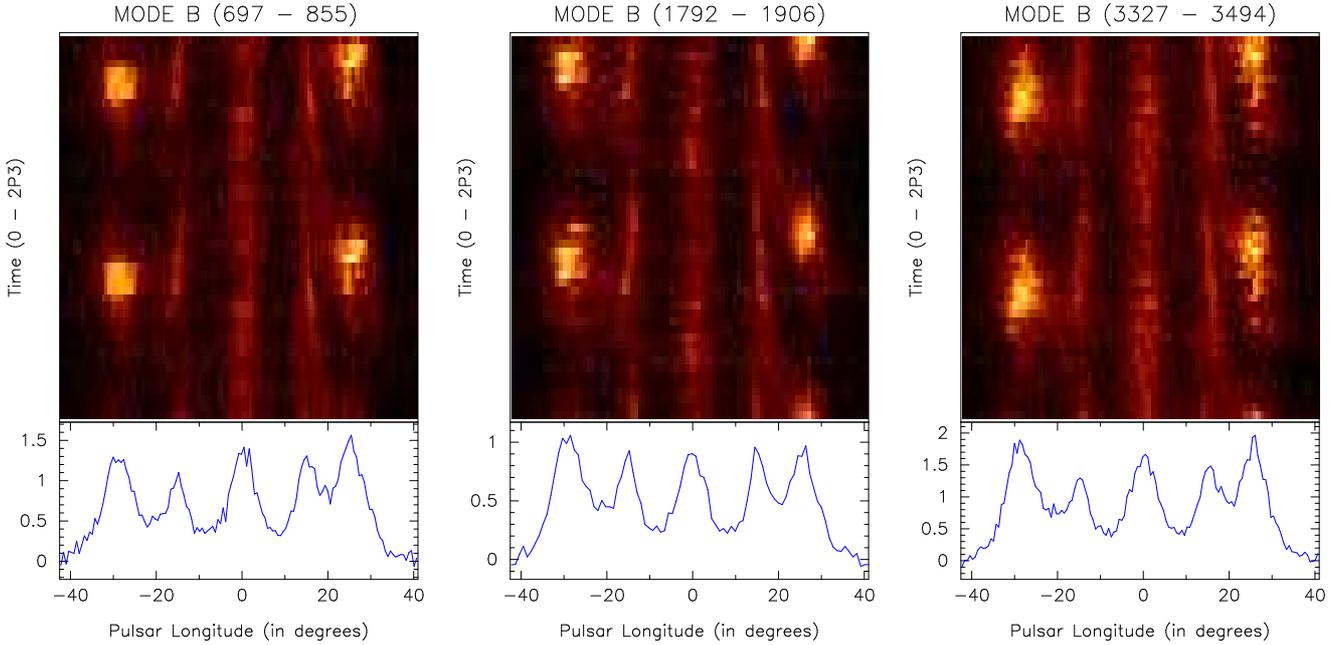

\begin{tabular}{@{}cr@{}cr@{}}
{\mbox{\includegraphics[scale=0.34,angle=0.]{ModeB_696-854_P3fold.ps}}} &
{\mbox{\includegraphics[scale=0.34,angle=0.]{ModeB_1792-1906_P3fold.ps}}} &
\hspace{10px}
{\mbox{\includegraphics[scale=0.34,angle=0.]{ModeB_3326-3493_P3fold.ps}}} \\
\end{tabular}
\caption{The figure shows $P_3$ folded profile during emission mode B  
for three different sections, between pulse 697 and 855 (left panel), between 
pulse 1793 and 1907 (middle panel) and from pulse 3327 to 3494 (right panel). 
In order to present a full outline of the subpulse tracks the same profile have
been stacked on top to give a 2$P_3$ range along the y-axis. The outer cones 
show bright spots indicating phase stationary drifting. The inner cones show 
extended tracks indicating large phase variations. The direction of these 
tracks have opposite sense which is suggestive of bi-drifting nature.} 
\label{fig_p3fold}
\end{figure*}

\begin{table}
\resizebox{\hsize}{!}{
\begin{minipage}{80mm}
\caption{Subpulse Drifting in PSR J2006$-$0807}
\centering
\begin{tabular}{c|@{\hskip3pt}c@{\hskip3pt}|@{\hskip3pt}c@{\hskip3pt}|cc|@{\hskip3pt}c@{\hskip3pt}|}
\hline
 Pulse & Mode & $f_p$ & FWHM & S & $P_3$ \\
   &  & ($cy/P$) & ($cy/P$) &  & ($P$) \\
\hline
   &  &  &  &  &  \\
 100 - 160 & A & 0.0342$\pm$0.0107 & 0.0252 & 24.0 & 29.3$\pm$9.2 \\
   &  &  &  &  &  \\
 616 - 674 & A & 0.0337$\pm$0.0136 & 0.0320 & 13.7 & 29.7$\pm$12.0 \\
   &  &  &  &  &  \\
 697 - 855 & B & 0.0667$\pm$0.0054 & 0.0128 & 40.5 & 15.0$\pm$1.2 \\
   &  &  &  &  &  \\
 1716 - 1793 & A & 0.0390$\pm$0.0071 & 0.0166 & 36.8 & 25.6$\pm$4.6 \\
   &  &  &  &  &  \\
 1793 - 1907 & B & 0.0699$\pm$0.0039 & 0.0092 & 60.8 & 14.3$\pm$0.8 \\
   &  &  &  &  &  \\
 2030 - 2123 & A & 0.0371$\pm$0.0107 & 0.0252 & 20.1 & 27.0$\pm$7.8 \\
   &  &  &  &  &  \\
 3233 - 3304 & A & 0.0338$\pm$0.0107 & 0.0253 & 20.7 & 29.6$\pm$9.4 \\
   &  &  &  &  &  \\
 3327 - 3494 & B & 0.0657$\pm$0.0035 & 0.0082 & 76.3 & 15.2$\pm$0.8 \\
   &  &  &  &  &  \\
 4188 - 4240 & B & 0.0675$\pm$0.0184 & 0.0434 & ~8.5 & 14.8$\pm$4.0 \\
   &  &  &  &  &  \\
 4702 - 4798 & A & 0.0367$\pm$0.0098 & 0.0231 & 20.2 & 27.3$\pm$7.3 \\
   &  &  &  &  &  \\
\hline
\end{tabular}
\label{tabdrift}
\end{minipage}
}
\end{table}

\noindent
We have carried out a detailed analysis of the subpulse drifting seen in the 
emission modes A and B. The drifting analysis was particularly challenging, 
firstly because of the mixture of drifting and non-drifting modes in the pulse 
sequence, and secondly because there was no clear boundary between the modes A 
and B. As mentioned earlier in section \ref{subsec:mod}, we have visually 
inspected the entire single pulse sequence to identify the modes A and B and 
investigated the nature of subpulse drifting. We have measured the LRFS of
individual pulse sequences in the two modes. Two typical examples of LRFS are 
shown in figure \ref{fig_lrfsdrift}, where the left panel corresponds to a 
sequence between pulse range 1716 and 1793~in mode A, and the right panel shows
the LRFS for a sequence in mode B between pulses 697 and 855. The peak 
frequency corresponding to drifting periodicity is seen primarily in the conal 
components, with the outer cones being most prominent. The core component 
either showed a low level intensity (mode A) or drifting was completely absent 
(mode B). In addition the phase variations corresponding to the peak frequency 
(top window in figure \ref{fig_lrfsdrift}) also show significant evolution 
across the different components. The figure also shows the errors in phase 
values which were computed as :\\\\
$\delta \phi= \frac{\displaystyle |x\delta y- y \delta x| }{\displaystyle x^2+y^2}; \quad$ where $\quad x=Re(I_\nu), \quad  y= Im(I_\nu)$, \\\\
where $\delta x$ and $\delta y$ have been computed from the baseline rms of the
real and imaginary parts of the spectral power at horizontal bins corresponding
to the peak drifting frequency ($f_p$). 

The phase behaviour provides a clear distinction between subpulse drifting and 
other periodic phenomenon like periodic nulling or amplitude modulation (see 
figure \ref{fig_ampmod}). In both emission modes the phases corresponding to 
the outer cones are relatively flat, which transitions to large variations for 
the inner cones. In order to further explore the phase variations we have 
carried out $P_3$ folding for all drifting sequences. In most of these cases 
the number of pulses were not adequate to form a stable profile. However, in a 
few cases, in mode B, we could establish the subpulse behaviour from the $P_3$ 
folded profiles. In figure \ref{fig_p3fold}, the $P_3$ folded profile for 
three sequences have been shown, pulses 697 and 855 (left panel), between pulse
1793 and 1907 (middle panel) and from pulse 3327 to 3494 (right panel). 
The outer cones are seen as bright spots which are indicative of their phase 
stationary nature. The inner cones on the other hand show extended tracks which
suggest large phase variations. Additionally, these tracks are in opposite 
directions with the second component showing a positive drifting and fourth 
component with negative drifting. This behaviour is also seen in the 
fluctuation spectra phase and is indicative of the rare phenomenon known as 
bi-drifting, where the subpulse drifting in different components have opposite 
directions \citep{cha05}.

\begin{table}
\resizebox{\hsize}{!}{
\begin{minipage}{80mm}
\caption{Slope of the drifting phase variations in the inner cones}
\centering
\begin{tabular}{ccccc}
\hline
 Pulse & Mode & \multicolumn{2}{c}{$d\phi/d\psi$ (\degr/\degr)} \\
   &  & COMP 2 & COMP 4 \\
\hline
   &  &  &  \\
 697 - 855 & B & -32.7$\pm$5.7 & 65.6$\pm$7.6 \\
   &  &  &  \\
 1716 - 1793 & A & -31.1$\pm$5.0 & 58.2$\pm$6.1 \\
   &  &  &  \\
 1793 - 1907 & B & -42.9$\pm$5.0 & 75.5$\pm$5.9 \\
   &  &  &  \\
 2030 - 2123 & A & -40.5$\pm$6.2 & 56.5$\pm$5.5 \\
   &  &  &  \\
 3233 - 3304 & A & -45.7$\pm$4.4 & 76.4$\pm$5.3 \\
   &  &  &  \\
 3327 - 3494 & B & -25.1$\pm$2.8 & 66.6$\pm$5.6 \\
   &  &  &  \\
\hline
\end{tabular}
\label{tabdriftphs}
\end{minipage}
}
\end{table}

\begin{figure}
\begin{center}
\includegraphics[scale=0.72]{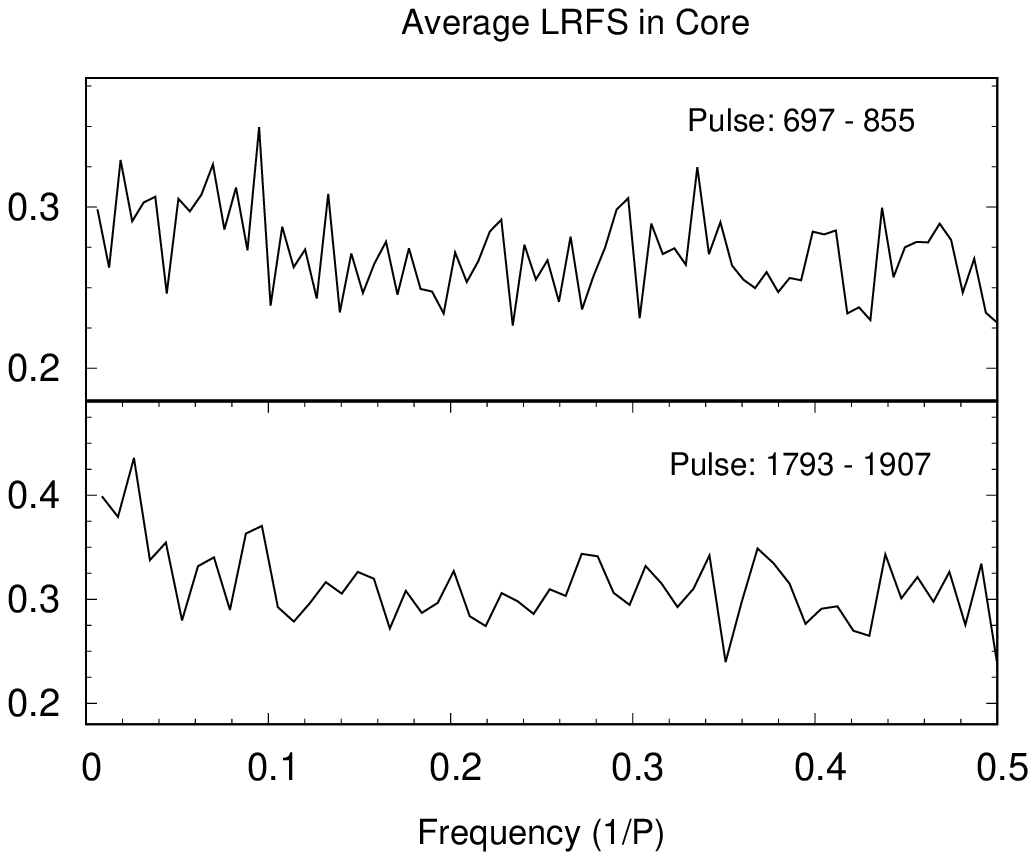}\\
\vspace{10px}
\includegraphics[scale=0.72]{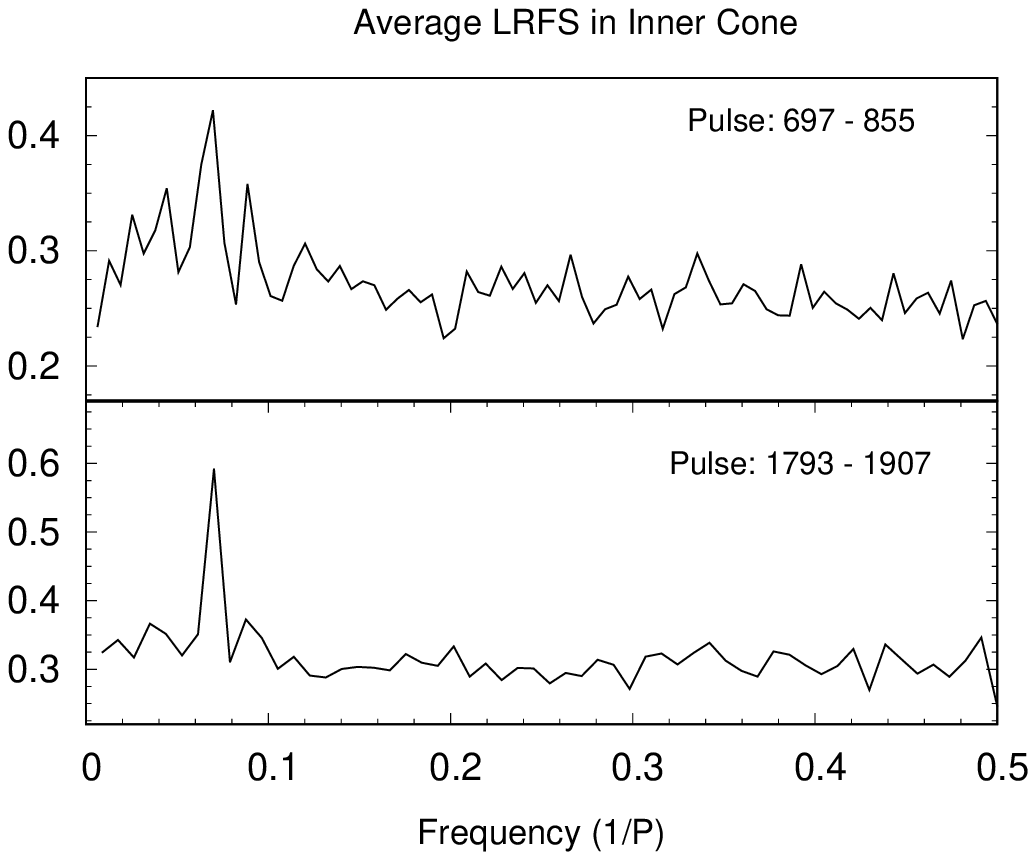}\\
\vspace{10px}
\includegraphics[scale=0.72]{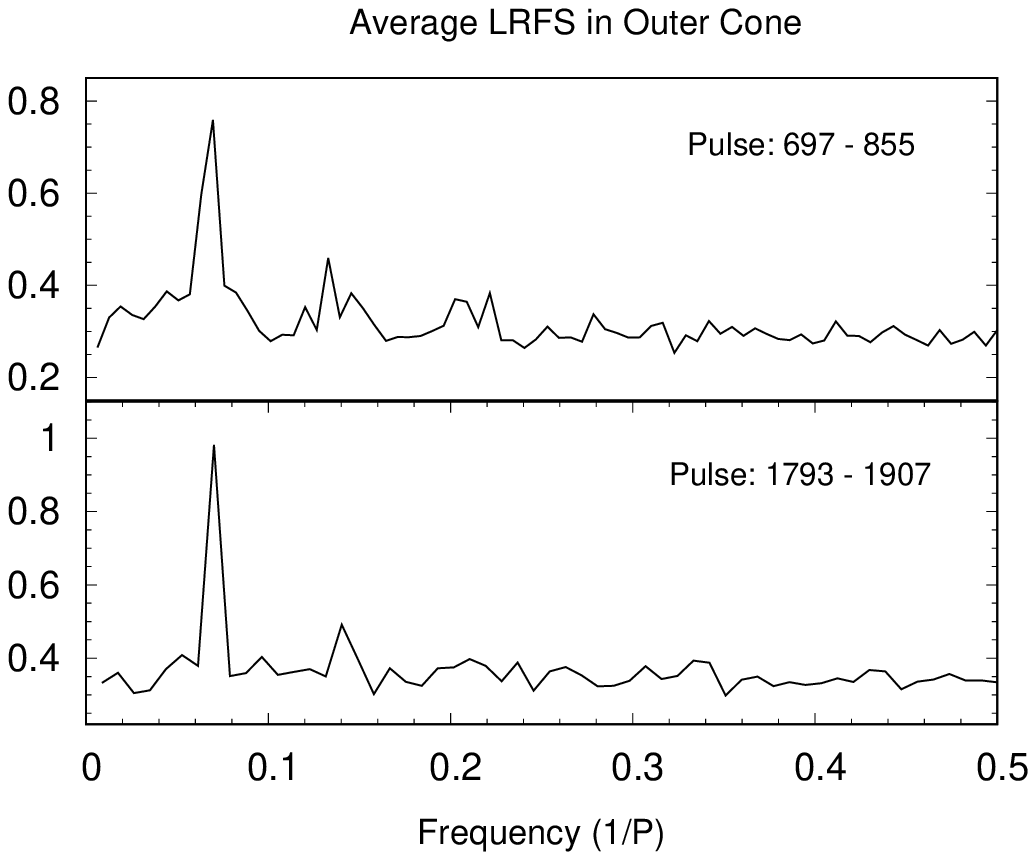}\\
\end{center}
\caption{The figure shows the Longitude Resolved Fluctuation Spectra (LRFS) 
averaged across the core (top panel), inner cone (middle window) and outer 
conal (bottom window) components, during two intervals in mode B. The top panel
corresponds to the pulses between 697 and 855, while the bottom panel 
corresponds to pulse range 1793 and 1907. The periodicity corresponding to 
drifting is clearly seen in the inner and outer conal components. However, no
clear drifting feature is visible in the core component.}
\label{fig_compfft}
\end{figure}

The drifting in modes A and B have different periodicities, with mode A showing
a much lower frequency peak compared to mode B in the fluctuation spectra. In 
Table \ref{tabdrift} we report all the sections in the pulse sequence where a 
clear drifting peak could be measured from the LRFS. In each case we have also 
estimated the S factor, which is defined as the ratio between peak height and 
Full Width at Half Maximum (FWHM) of the feature, and gives a measure of the 
peak strength. The Table lists the pulse range corresponding to each 
measurement along with emission mode, $f_p$, FWHM of each feature, S factor, 
and drifting periodicity ($P_3$) in units of rotation period ($P$). The 
error in the peak frequency is estimated as FWHM/2$\sqrt{2\ln(2)}$, where 
2$\sqrt{2\ln(2)}$ is the scaling for gaussian approximation \citep{bas16}. The 
Table reports six pulse sequences in mode A with a mean $P_3$ of 
27.1$\pm$8.9~$P$. Additionally, four sequences in mode B are also reported with
a mean $P_3$ value of 14.8$\pm$0.3~$P$. Within measurement errors the 
drifting periodicity during mode A is roughly two times that of mode B. In 
order to explore any harmonic relationship between the two we have investigated
the phase behaviour of the inner conal components for the different sequences
where the phase variations could be clearly distinguished in the LRFS. A linear 
approximation to the phase variations for each sequence was estimated and the 
phase gradient ($d\phi/d\psi$) is reported in table \ref{tabdriftphs}. The 
gradients in the two conal components are of opposite signs consistent with the
bi-drifting behaviour and the magnitude is higher for the fourth component 
compared to the second component. If the periodicities during the two modes are
harmonically related then for each component the phase gradient during mode B 
(first harmonic) should be a factor of 2 greater than the gradient in mode A. 
No, such dependence is seen in the phase behaviour during the two modes which 
rules out any harmonic dependence between the two.

We have also investigated the nature of subpulse drifting in the 
individual components and whether any low level drifting is seen in the core 
component. In figure \ref{fig_compfft} we show the average LRFS for the core 
(top window), inner cones (middle window) and the outer cones (bottom window), 
for two pulse ranges between 697 and 855 (top panel in each window) and between
pulses 1793 and 1907 (bottom panel), when the pulsar was in mode B. The 
presence of prominent drifting peaks is seen for the inner and outer cones. 
However, no such features are seen for the core component demonstrating the 
absence of drifting in the core. However, there is a possible low level, wide 
structure present at the lower frequencies. This is likely due to the presence 
of periodic amplitude modulations in these modes in addition to drifting. 
However, these structures cannot be related to the sharp low frequency features
seen in modes C and D in section \ref{subsec:null} as a result of nulling, 
since no periodic nulling is seen in during these modes.

\section{Discussion}\label{sec:disc}
\subsection{The interaction between Nulling and Mode changing}
\noindent
The single pulses in PSR J2006$-$0807 show large diversity with four distinct 
emission modes, two of which show subpulse drifting, in addition to long 
intervals of nulling. There are only a few known pulsars which exhibit more 
than two emission modes along with subpulse drifting. The most well studied 
pulsar in this group is PSR B0031$-$07 with three distinct drifting modes 
\citep{viv97,smi05,mcs17}. The pulsar also shows the presence of long duration 
periodic nulls \citep{bas17}. It has a conal profile and belongs to the S$_d$ 
profile class. Other examples include the pulsars B1819$-$22 \citep{bas18b}, 
B1918+19 \citep{ran13}, J1944+17 \citep{klo10} and B2319+60 \citep{wri81}. The 
pulsar B1819$-$22 has a conal double profile and exhibit two distinct drifting 
modes in addition to a non-drifting state. The pulsar also shows short duration
nulls during all emission states which are periodic in nature. In case of PSR 
B1918+19 there are three distinct drift modes and an additional disordered 
mode. The pulsar also shows the presence of nulling which is periodic. 
The pulsar has a conal profile belonging to the $_c$T class. The pulsar 
B1944+17 also has a $_c$T profile and nulls for around 70\% of the time. In the
burst state the pulsar shows the presence of four modes, three of which show 
subpulse drifting. The nulling can be divided into two categories, short 
duration nulls in between the burst states which likely showed low level 
emission, and long duration nulls lasting up to 300 periods. The presence of 
periodicity was detected associated with the long duration nulls \citep{bas17}.
Finally, the pulsar B2319+60 shows the presence of three drifting modes and 
nulls for roughly 30\% of the time. There is no study demonstrating the 
periodicity of the nulls in this case, however, the fluctuation spectra shows 
the presence of a low frequency feature which maybe associated with nulling. 
The pulsar profile has four conal components and belongs to the $_c$Q class. 
All the previously reported pulsars showing multiple mode changing along 
with subpulse drifting are associated with conal profiles. On the contrary the
pulsar J2006$-$0807 has five components, including a central core component, 
and belongs to the M profile class. This is the first case where we have found 
multiple mode changing, drifting and nulling in the presence of the core 
emission. As a result we can conclude that such phenomenon are not restricted 
to conal pulsars alone and no preferred line of sight geometrical configuration
is responsible for such systems.

There is no clear boundary of transition between the two drifting modes A and B
which is also unique to this pulsar. A gradual shift in the drifting 
periodicity was reported for the pulsar B0943+10 \citep{bac11}. The variation
was small and seen in the Burst mode as an exponential decay over timescale of 
few hours. There was no separate mode identification carried out for this 
evolution in drifting. In the case of PSR B0809+74 which shows systematic 
drifting, a null associated mode changing has also been reported \citep{van02}.
The drift rate shows a gradual variation lasting several hundred periods after
certain intervals of nulls, before transitioning to the earlier mode. There is 
no clear boundary during the mode transition which lasts for around 20 $P$. 
Similarly, the evolution between modes A and B in PSR J2006$-$0807 occurs over 
10-20 $P$ with distinct evolution in drifting periodicity as well as profile 
shape. The pulsar also shows the presence of two types of nulls, the short 
duration nulls in between the burst states in the non-drifting modes, as well 
as longer duration nulls lasting up to 100 $P$. This behaviour is different from
the other cases where the short nulls are present between both the drifting and
non-drifting modes. Additionally, only the short duration nulls in the 
non-drifting modes show clear periodicity. This clearly distinguishes the two 
phenomenon which was not apparent in previous cases. The presence of periodic 
nulling is reported in around twenty pulsars \citep{her09,bas17}. In this group
there are seven pulsars which show presence of both mode changing and periodic 
nulling, including the the six pulsars discussed in this section. In recent 
works \citep{bas16,mit17,bas17,bas18a,bas18b} a clear distinction has been 
between the different periodic phenomenon in pulsars. The subpulse drifting has
different physical properties compared to periodic nulling and periodic 
amplitude modulations. This is highlighted by the dependence of the respective 
periodicities with $\dot{E}$. Additionally, it has also seen that the subpulse 
drifting is strictly a conal phenomenon while the nulling and mode changing 
affects all the components, both core and cones, at the same time. 

\subsection{Periodic Nulling in presence of Core component}
\noindent
As discussed above the pulsar shows the presence of periodicity associated with
nulling as well as subpulse drifting. This is the eighth case where the 
presence of subpulse drifting and periodic nulling were seen in the same system 
\citep{bas17,bas18b}. However, this is the first time that we have clearly seen
periodic nulling in the presence of the core component, where the core and cone
nulls simultaneously. Our analysis provides conclusive proof that the periodic 
nulling is a separate phenomenon, completely different from subpulse drifting. 
Firstly, the two phenomena were seen in different emission modes with 
contrasting single pulse behaviour. The subpulse drifting was seen in the 
emission modes A and B where no significant nulling was seen, while the 
periodic nulling was observed in the modes C and D where no systematic drift 
pattern was visible. Secondly, the drifting was seen as subpulse variation 
affecting each component differently, while the periodic nulling affected the 
entire single pulse, with either the presence or absence of emission across the
pulse window. This was also reflected in the phase behaviour, with the phases 
being completely flat across the pulse window during periodic nulling, while 
they showed significant structure during drifting. Finally, the behaviour of 
the core emission clearly demonstrated the difference between the two. While 
the core and the cones nulled simultaneously during periodic nulling, the 
subpulse drifting was absent in the core component. The two phenomenon have 
been understood in the past \citep{her09} in terms of the carousel model 
\citep{rud75,gil00,des01}, where the emission beam was expected to consist of 
circulating beamlets which were responsible for subpulse drifting. The longer 
periodicity associated with periodic nulling was interpreted as a circulation 
time, $P_4$, of each beamlet around the emission beam. The periodic nulls were 
either LOS traverse between empty regions of the circulating pattern or 
corresponded to missing or weak beamlets \citep{ran13}. \cite{bas16,bas17} 
found that subpulse drifting was seen in pulsars with spindown energy loss 
($\dot{E}$) $<$ 10$^{33}$ erg~s$^{-1}$, while no such restriction appeared for 
periodic nulling and low frequency amplitude modulations. Additionally, $P_3$ 
was found to be inversely correlated with $\dot{E}$, while no such dependence 
was seen for $P_N$. In a more recent work investigating subpulse drifting, 
\citet{bas19} established subpulse drifting to be primarily a conal 
phenomenon, while periodic amplitude modulation and periodic nulling were 
associated with the entire pulse window. As a result these studies claimed 
different physical processes to be responsible for subpulse drifting and 
periodic nulling. Our analysis for PSR J2006$-$0807, presented in this work, 
provides further and more conclusive evidence to support this claim.

\subsection{Understanding the Drifting Behaviour}
\noindent
One of the most striking feature about the pulsar's single pulse behaviour is 
the nature of subpulse drifting seen across the pulse window. The central line
of sight traverse of the emission beam provides a wide window into the drifting 
process and have given us improved insights into the inner cones of a M type 
profile class. The fluctuation spectral studies of drifting give measurements 
of drifting periodicity as well as phase variations corresponding to subpulse 
motion across each component in the pulsar profile. 
As reported in section \ref{sec:drift} the phase variations are very different 
for the inner and outer cones. The outer cones were mostly phase stationary 
with less than 50\degr~variations across each component. On the other hand the 
inner cones, particularly in mode B, showed large variations in excess of 
100\degr~across each component. A recent survey of subpulse drifting was 
conducted by \citet{bas19}, where the phenomenon, seen in around 60 pulsars, 
was classified into four groups based on their phase behaviour. A connection 
between the phase variations and profile class, and as an extension emission 
geometry, was seen. The drifting in pulsars with a central core component were 
classified as `Low-mixed' phase modulated drifting. It was previously expected 
that the subpulse drifting in central LOS traverse, associated with core 
emission, was phase stationary \citep{ran86}. However, \citet{bas19} noted 
that certain pulsars, like PSR B1237+25, had indications of considerable phase 
variations, particularly for the inner conal components. The indications 
for such phase behaviours in PSR B1237+25 was also seen in \citet{ran03}. No 
detailed drifting phase measurements were available for PSR J2006$-$0807, but 
the pulsar was also classified in the `Low-mixed' category. The phase behaviour
in both modes A and B, reported here, is consistent with this classification. 
To the best of our knowledge this is the first pulsar where we have 
conclusively found large phase variations in a system with a core component. It
is possible that the phase behaviour seen for the inner and outer cones are 
more common in the M type profiles, as also indicated in B1237+25 \citep[see 
the phase plots in ][]{bas19}. Detailed observations exploring the drifting 
phase behaviour in other such pulsars is required to explore this further. 

In addition to large variations the phases also show opposite directions in the
inner cones resulting in bi-drifting. This is only the fourth known example of 
bi-drifting phenomenon, with the other three being J0815+0939 \citep{cha05,
sza17}, B1839$-$04 \citep{wel16} and J1034$-$3224 \citep{bas18a}, which are all
wide profile conal only pulsars. In PSR J0815+0939 there are four components 
in the profile which belongs to the $_c$Q class. The second component shows 
positive drifting while the other three has negative drifting. In PSR 
B1839$-$04 the profile has two components with opposite drifting directions. 
The pulsar J1034$-$3224 has four distinct conal components in addition to a 
precursor like emission \citep{bas15}. The phase variations are opposite in 
every alternate conal component of this pulsar. 
The bi-drifting phenomenon is difficult to understand using standard model
of plasma generation in Inner Acceleration Region \citep[IAR,][RS75]{rud75}. An
usual circulation of sparks around the magnetic axis in a strictly dipolar 
polar cap cannot give drifting in opposite directions for any line of sight 
traverse of the emission beam. A number of models for a modified IAR has been 
proposed to explain the bi-drifting phenomenon. \citet{qia04a,qia04b} suggested
the IAR to consist of an inner annular gap inside the traditional RS75 gap. The
drift velocity has opposite directions in the two gaps which was used to 
explain the bi-drifting effect. However, as pointed out the drawback of this 
model is the requirement of different number of rotating sparks circulating at 
varying speeds in the two gaps to reproduce the observed drifting. An empirical
model was proposed by \citet{wri17} where the emission beam was assumed to be 
elliptical and tilted with respect to the fiducial plane. It was shown that 
specialized line of sight traverses can sample the circulating pattern in a 
manner that the drifting in different components are reversed.
A more physical model was suggested by \citet{sza17}, where the IAR was assumed
to be dominated by non-dipolar fields. It was suggested that the sparks 
circulated around a point of maximum potential within the polar cap. This 
configuration reproduced the elliptical beams suggested by \citet{wri17}, 
thereby explaining the bi-drifting phenomenon. The phase behaviour in PSR 
J2006$-$0807 is different primarily due to the presence of the core component 
which does not drift, and bi-drifting is only seen in the inner cones. This 
configuration also greatly constrains the line of sight geometry which is 
supposed to traverse the emission beam centrally. In addition to the 
bi-drifting in the inner cones, the outer cones exhibit almost phase stationary
behaviour. Some phase variations are expected for pulsars with lower 
magnetic inclination angles even for central line of sight traverse. However, 
the difference in the phase behaviours of the inner and outer cones cannot 
simply be explained by a low inclination angle.
In contrast to the widely varying phase behaviour across the pulse window the
peak frequency corresponding to the drifting periodicity remains constant for
all components. This implies that the drifting speed remains constant across 
the pulse window, despite the direction of motion having large variations as 
indicated by the phase behaviour. In addition to the pulsar J2006$-$0807, 
\citet{bas19} have also found that the drifting periodicity remains constant 
across the pulse window for all pulsars with systematic drift patterns, despite 
large phase variations. 

\section{Summary}\label{sec:sum}
\noindent
We have carried out a detailed single pulse analysis of the five component 
pulsar J2006$-$0807. The pulsar profile was characterised by a central core 
component surrounded by pairs of inner and outer cones. The single pulses 
revealed the presence of four distinct emission modes along with intervals of 
nulling. In two emission modes, A and B, the pulsar showed the presence of 
subpulse drifting, while the other two modes, C and D, did not drift, but were
interspersed with periodic nulls. The pulsar spent roughly a third of its time
in the two drifting modes, another third in the two non-drifting modes and the 
remaining in the nulling state. We have estimated the average polarization 
properties and emission geometry, which suggested that the emission location
within the magnetosphere was unchanged for the different modes. Additionally, 
the polarization also indicated the presence of orthogonal moding which was 
connected with the mode changing phenomenon. The subpulse drifting showed 
diversity in the drifting behaviour across the different components, with the 
outer cones showing phase stationary drifting, the inner cones showing large 
phase modulated bi-drifting. The emission showed a gradual decrease in 
intensity as the pulsar transitioned from mode A to B. In addition, the 
fluctuation spectra also showed a low level wide structure at low frequencies 
during this emission state, primarily in the core component which was due to 
periodic amplitude modulation. The amplitude modulation was also likely to be 
present in the conal components but was hidden under the prominent drifting
feature.

\section*{Acknowledgments}
We thank the referee for the comments which helped to improve the paper. DM 
acknowledges funding from the grant ``Indo-French Centre for the Promotion of 
Advanced Research - CEFIPRA". We thank the staff of the GMRT who have made 
these observations possible. The GMRT is run by the National Centre for Radio 
Astrophysics of the Tata Institute of Fundamental Research.

\end{document}